\RequirePackage{fix-cm}
\documentclass[natbib,smallextended]{svjour3}       % onecolumn (second format)
\smartqed  % flush right qed marks, e.g. at end of proof
\usepackage{graphicx}
\usepackage{mathptmx}      % use Times fonts if available on your TeX
\usepackage{amssymb}
\usepackage{epstopdf}
\usepackage{times}
 \journalname{my journal}
\newcommand{\aap}{{Astron. Astrophys.}}
\newcommand{\apj}{{Astrophys. J.}}
\newcommand{\apjs}{{Astrophys. J. Supp.}}
\newcommand{\aj}{{Astronom. J.}}

\newcommand{\mnras}{Mon. Not. Roy. Astr. Soc.}
\newcommand{\ssr}{Space Sci. Rev.}
\newcommand{\nat}{Nature}
\newcommand{\pasj}{Proc. Astr. Soc. Jap.}

\begin{document}

\title{Disks and Jets}\thanks{Supported by ISSI, NSF}

\subtitle{Gravity, Rotation and Magnetic Fields}

%\titlerunning{Short form of title}        % if too long for running head

\author{John F. Hawley        \and
        Christian Fendt \and
        Martin Hardcastle  \and
        Elena Nokhrina  \and
        Alexander Tchekhovskoy\footnote{Einstein Fellow}
}

%\authorrunning{Short form of author list} % if too long for running head

\institute{John F. Hawley \at
              Department of Astronomy, University of Virginia \\
              \email{jh8h@virginia.edu}           %  \\
%             \emph{Present address:} of F. Author  %  if needed
           \and
          Christian Fendt \at
             Max Planck Institute for Astronomy,
             K\"onigstuhl 17,
             D-69117 Heidelberg,
             Germany \\
             \email{fendt@mpia.de}
          \and
           Martin Hardcastle \at
             School of Physics, Astronomy and Mathematics,
             U. Hertfordshire,
             Hatfield, UK\\
             \email{m.j.hardcastle@herts.ac.uk}
\and
           Elena Nokhrina \at
             Moscow Institute of Physics and Technology, Russia\\
             \email{nokhrinaelena@gmail.com}
\and
           Alexander Tchekhovskoy \at
             Department of Physics and Department of Astronomy,
             UC Berkeley,
             Berkeley, CA\\
             \email{atchekho@berkeley.edu}
}

\date{Received: date / Accepted: date}
% The correct dates will be entered by the editor

\maketitle

\begin{abstract}
Magnetic fields are fundamental to the dynamics of both accretion
disks and the jets that they often drive. We review the basic physics
of these phenomena, the past and current efforts to model them
numerically with an emphasis on the jet-disk connection, and the observational constraints on the role of magnetic
fields in the jets of active galaxies on all scales.
% \PACS{PACS code1 \and PACS code2 \and more}
% \subclass{MSC code1 \and MSC code2 \and more}
\end{abstract}

\section{Introduction}
\label{sec:intro}

%  JH

The topic of this workshop was the diverse astrophysical phenomena
enabled by the strongest magnetic fields in the Universe.  Gamma
ray bursts and magnetars provide two examples of extreme phenomena
where extraordinarily strong fields are manifest.  However, the
concept of ``strong'' is relative.  A magnetic field may be weak
in the sense that the energy density associated with the field is
small compared to other energies, e.g., thermal, rotational,
gravitational.  Such ``weak'' magnetic fields can, however,
enable phenomena that otherwise would not be possible.  In this
contribution we consider just such an example, namely accretion
disks and the jets they produce.  Magnetic fields are fundamental
to both accretion and outflows, and while in some specific cases
the magnetic fields may be strong in the conventional sense of total
energetics, for the most part the fields are likely to be relatively
weak by those same conventional metrics.  Nevertheless, magnetic
fields play the central dynamic role.

Gravity is the dominant force in creating the structure in the
universe.  When gravity is combined with angular momentum, the
result is a disk.  Since some angular momentum will inevitably be
present, disks are arguably the most basic type of object in the
universe.  Other structures, such as stars, result from disks that
have somehow shed most of their angular momentum.  Accretion is
also a major astrophysical power source.  As Lynden-Bell observed
\citep{LyndenBell69}, the gravitational energy available in black
hole accretion greatly exceeds that available through nuclear
reactions, with efficiencies anywhere from 6--40\% $mc^2$, versus
less than 1\% for nuclear fusion.

{\it Direct} observational evidence for magnetic fields in accretion
disks is very limited (measurements exist, e.g., for FU Orionis, 
\citealt{2005Natur.438..466D}). 
The strongest case for magnetic fields is
theoretical.  We know from observations that disks accrete at rates
that require considerable internal stress to transport angular
momentum.  As we shall discuss, this requires magnetic fields.

% CF 

{\it Indirect} evidence for magnetic fields in accreting systems comes
from observations of highly collimated outflows, or jets.  Jets are
one of the most striking signatures resulting from accretion,  and
are seemingly ubiquitous; they are associated with young stars,
micro-quasars or X-ray binaries (MQs, XRBs), and active galactic
nuclei (AGN).  The current understanding of jet formation is that
outflows are launched by {\em magnetohydrodynamic} (MHD) processes
in the close vicinity of the central object \citep{BP82,
1983ApJ...274..677P, 2007prpl.conf..277P, 2007prpl.conf..261S} It
is furthermore clear that accretion and ejection are related to
each other.  Jets may be a consequence of accretion, but accretion
may, in turn, be a consequence of a jet.  Accretion occurs when gas
loses some of its angular momentum, and one very efficient way to
remove angular momentum from a disk is to eject it vertically into
a jet or outflow \citep{1993ApJ...410..218W, 1993A&A...276..625F,
1995ApJ...444..848L, 1997A&A...319..340F}.  Not all details of the
physical processes involved are completely understood yet.

In stellar sources, a central stellar magnetic field is surrounded
by a disk carrying its own magnetic flux.  Such a geometrical setup
can be found in young stars, cataclysmic variables, high-mass and
low-mass X-ray binaries, and other micro-quasar systems.  Systems
such as AGN or $\mu$-quasars have a black hole at their center, and
a black hole cannot have an intrinsic magnetic field.  The surrounding
accretion disk, however, can support the magnetic flux needed for
jet launching.  In this case, the interaction of the black hole
itself with the ambient field may launch a highly magnetized Poynting
flux-driven axial flow \citep{BZ-77}.

Although jets are found in a wide range of systems, from protostars to
AGN, from the observational point of view protostellar and
relativistic jets are quite different. For protostars we can measure
Doppler-shifted forbidden line emission to obtain gas velocities,
densities or temperatures, and these provide a good estimate on the
mass flux of the outflow. Evidence concerning the magnetic field,
however, is mainly indirect: magnetic flares of large-scale
protostellar stellar magnetic fields indicate that protostellar jets
originating from a area close to the star, originate in an
magnetized environment. As noted above, in a few objects there
  are direct measurements of the strength or orientation of magnetic
  fields in the disks \citep[e.g.][]{2005Natur.438..466D,Stephens+14}.
Concerning the jet magnetic field, direct evidence is less clear. So
far an estimate is available for only one source, HH80/81, obtained
from observed synchrotron emission \citep{2010Sci...330.1209C}. The
situation is completely reversed for relativistic jets. While there is
plenty of information concerning the jet magnetic field structure, and
some constraints on field strengths, speeds are harder to measure,
there is no direct measure of the mass fluxes involved,  and the
  magnetic properties of the accretion flow are not directly
  observable. We do not even know for certain whether the matter
content of these jets is leptonic or hadronic (Section \ref{sec:obs}).

While protostars and AGN are the most common jet sources known,
  the typical characteristics of jet systems --- the existence of an
  accretion disk and a strong magnetic field --- are also present in
  other astrophysical sources, namely in cataclysmic variables, high
  mass or low mass X-ray binaries, and pulsars. Resolved Chandra
  imaging of the Crab \citep{2002ApJ...577L..49H} and the Vela pulsar
  \citep{2003ApJ...591.1157P} do indeed show elongated, highly
  time-variable but persistent structures emerging along the
  rotational axis of these systems. Whether these structures are
  similar to the jets observed in protostars and AGN or are features
  intrinsic to the pulsar wind nebula is not yet clear. {\it
    Persistent} jets are not widely observed in cataclysmic variables
  (either in the highly transient dwarf novae or the more stable
  nova-like objects), although there are reports of unresolved,
  steep-spectrum radio emission, consistent with jet activity, in both
  types of system \citep{2008Sci...320.1318K,2011MNRAS.418L.129K}. It
  is not clear why jets are not more commonly observed in these
  systems, although some jet launching models suggesting explanations
  exist in the literature \citep{2004A&A...422.1039S}. One basic
  reason may be the lack of axisymmetry in cataclysmic variables, as
  this is thought to be another major condition to launch a jet for a
  considerable period of time \citep{1998A&A...334..750F}. Alternatively,
  the observability of jet-related synchrotron emission depends
  strongly on the local environment of the jet, and it is possible
  that the conditions for this emission to be observed are simply not
  typically met in CVs.

% JH

In this workshop proceeding we will review the many fundamental
roles played by magnetic fields in accretion and outflows, both
from a theoretical and an observational point of view.
The principal processes involved in jet formation can be summarized
as follows \citep{BP82, 1983ApJ...274..677P,
2007prpl.conf..277P, 2007prpl.conf..261S}, 

\begin{itemize} 

\item[(1)] Jets are powered by gravitational energy released through
accretion and by rotational energy of the disk and/or the central
star (or black hole).  Magnetic flux is provided by the star-disk (or black hole-disk)
system, possibly by a disk or stellar dynamo, or by the advection of
the interstellar field.  The star-disk system also drives an electric
current.

\item[(2)] Accreting plasma is diverted and launched as a plasma
wind (from the stellar or disk surface) coupled to the magnetic
field and accelerated magneto-centrifugally.

\item[(3)] Inertial forces wind up the poloidal field inducing a
toroidal component.

\item[(4)] The jet plasma is accelerated magnetically (conversion
of Poynting flux).

\item[(5)] The toroidal field tension collimates the outflow into
a high-speed jet beam.

\item[(6)] The plasma velocities subsequently exceed the speed of
the magnetosonic waves.  The super-fast magnetosonic regime is
causally decoupled from the surrounding medium.

\item[(7)] Where the outflow meets the ISM, a shock develops,
thermalizing the jet energy.

\end{itemize}

\begin{figure}[t!]
\includegraphics[width=\textwidth,clip=true]{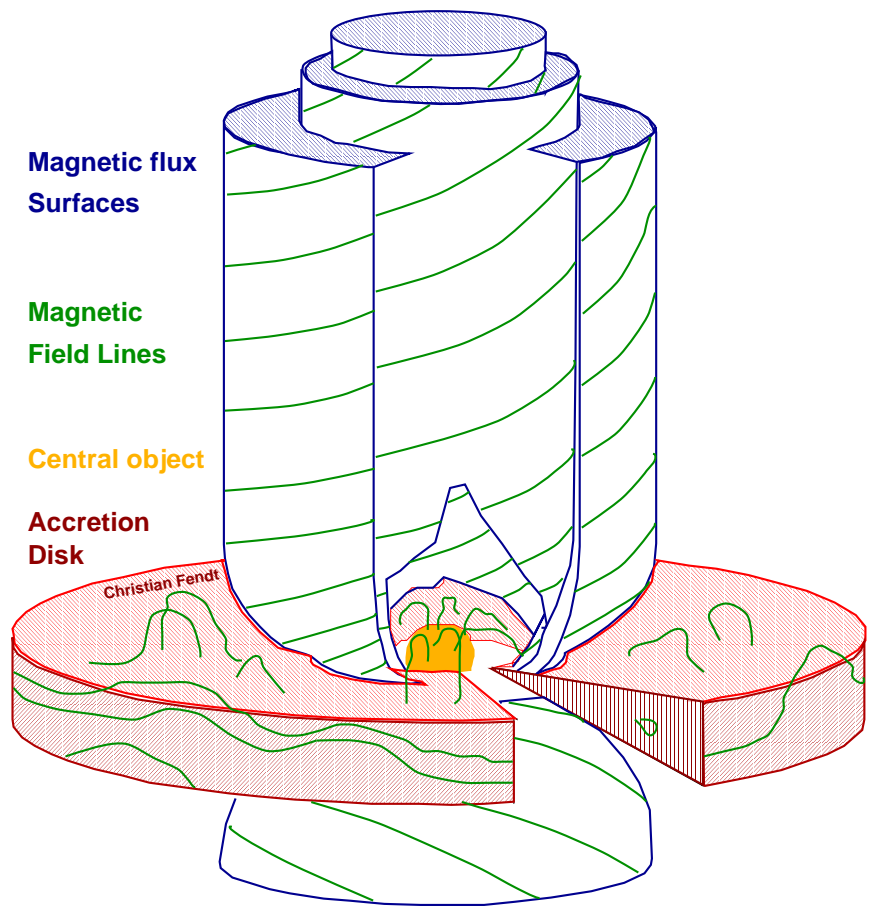}
\caption{Model sketch of MHD jet formation, indicating the six
generic problems to be solved: launching, acceleration, and
collimation; disk structure; magnetic field origin; asymptotic
interaction; central spine jet.
}
\label{fig:model}
\end{figure}

The overall process as outlined above is complex and takes place
over a wide range of both length- and time-scales.  Due to this
complexity, the aspects of the jet problem have to be tackled
independently.  One may distinguish six principal topics
(Fig.~\ref{fig:model}), roughly corresponding to the stages in the
overall picture described above.  These include:

\begin{itemize}

\item[(1)]
 The accretion disk structure and its evolution, including thermal effects 
 and the origin of turbulence.

\item[(2)]
 The origin of the jet magnetic field, possibly through a disk dynamo, a stellar
 dynamo, or by the advection of ambient magnetic field.

\item[(3)]
 The ejection of disk material into wind, thus the transition from accretion
 to ejection.

\item[(4)]
 The collimation and acceleration of ejected material into jets.

\item[(5)]
 The propagation of the asymptotic jet, its stability and
 interaction with ambient medium. A related question is the feedback
 of jets on star or galaxy formation.

\item[(6)]
 The possible presence and impact of a central spine jet, e.g. a stellar wind 
 or black hole jet, in comparison to a jet originating with the disk. Under what
 circumstances are disk jets and spine jets present or absent?

\end{itemize}

In this article we begin by describing the theoretical support for the
role of magnetic fields in the accretion disk itself (Section
\ref{sec:1}). We then discuss the modeling of magnetically driven
jets (Section \ref{sec:jets}) and their relationship to the disk (Section
\ref{sec:accretion-jet}). Observational constraints on field
properties and their relation to models are discussed in Section
\ref{sec:obs}. Our conclusions are presented in Section \ref{sec:conclusion}.

\section{Magnetic Fields and Accretion Disks}
\label{sec:1}

We begin with the accretion disk itself, and the fundamental role
played by magnetic fields.  Throughout the history of disk theory,
``shedding the angular momentum'' has been easier said than done.
The absence of an obvious mechanism to do so was considered a serious
objection to Laplace's disk hypothesis for solar system formation,
for example.  By the time that the basic theory of accretion was
laid out by \cite{Shakura73} and \cite{Lynden74} it was accepted
that Nature accomplished the transport of angular momentum by some
means, though that means was unknown to astronomers.  Shakura and
Sunyaev introduced the ``alpha viscosity'' parameterization as a
means of sidestepping the uncertainty.  In this model the internal
stress is proportional to the total pressure, $T_{R\phi} = \alpha P$.

The physical viscosity in the gas within an accretion disk is known
to be far too small to account for observed accretion rates.  If
the disk were turbulent, however, the Reynolds stresses associated
with that turbulence could transport angular momentum at the necessary
rate.  This seemed like a straightforward solution to the problem.
Because the viscosity is so small, the Reynolds number (ratio of
characteristic velocity times a characteristic length divided by
the viscosity) of the gas is correspondingly huge, and in terrestrial
contexts high Reynolds number flows are generally turbulent.  But
gas in Keplerian orbits is stable to perturbations by the Rayleigh
criterion, which simply requires that angular momentum increase
with radius, $dL/dR > 0$.  The positive epicyclic frequency associated
with Keplerian orbital flows is strongly stabilizing, and it is not
energetically favorable for turbulence to develop and be sustained
from the background angular momentum distribution \citep{BHS96,BH98}.

The linear stability of Keplerian flows was, of course, recognized
early on as a difficulty for the creation of turbulence in disks.
However, it was known that in the laboratory simple shear flows
display {\it nonlinear} instability. \cite{BHS96} argued that this
nonlinear instability represented the boundary between the
Rayleigh-stable and Rayleigh-unstable regimes; when the epicyclic
frequency goes to zero the linear response vanishes, leaving nonlinear
terms as the lowest order effect. Given the importance of the question
of the nonlinear stability of hydrodynamic Keplerian flow, several
groups have carried out fluid experiments. \cite{Ji06} examined the
stability of a Rayleigh-stable Couette flow and found no evidence of
significant turbulence. \cite{Paoletti11}, on the other hand, found a
breakdown into turbulence in their experiment. The experiments of
\cite{Schartman12} found no transition to turbulence, and attributed
the \cite{Paoletti11} results to turblence generated at the end caps
of the Couette cylinder, a conclusion further supported by the
numerical simulations of \cite{Avila12}. More recent laboratory
experiments by \cite{Edlund14} are also consistent with stability.
While work will no doubt continue on this important issue, at the
moment the weight of the theoretical and experimental data lie with
the nonlinear hydrodynamic stability of Keplerian flow.

\subsection{Basic MRI Physics}
\label{sec:2}

It was, of course, recognized early on that magnetic fields could
transport angular momentum through Maxwell stresses, $-B_r B_\phi/4\pi$,
but unless the magnetic pressure were near the thermal pressure,
it was expected that the effective $\alpha$ would be relatively
small.  As it turns out, magnetic fields render the disk linearly unstable
to the \textit{magneto-rotational instability} \citep[MRI]{BH91}
even when (and, indeed, \textit{only} when) the magnetic field is
weak.  The MRI can be visualized intuitively in the form of a spring
connecting two orbiting masses \citep{BH98}, one in a slightly lower
orbit than the other.  Because angular velocity decreases with $R$,
$\Omega \propto R^{-3/2}$, the inner mass pulls ahead of the outer
mass.  This causes the spring to stretch; the tension force pulls
back on the inner mass, and pulls forward on the outer.  The effect
of this is to transfer some angular momentum from the inner to the
outer mass, but in doing so the inner mass drops to a lower orbit,
increasing its angular velocity, increasing the relative separation
of the masses and increasing the spring tension force.  The process
runs away unless the spring is strong enough to force the masses
to remain together, which happens only if the tension exceeds the
orbital dynamical force.  If we replace the spring with a magnetic
field, the spring tension is replaced with the magnetic tension
associated with the Maxwell stress.  In a magnetized disk the linear
stability requirement is no longer the Rayleigh criterion, but is
instead a requirement that angular {\it velocity} increase outward,
$d\Omega/dR > 0$.  A Keplerian disk is unstable.

There are several remarkable properties of the MRI. One is that the
question of stability does not depend on the magnetic field strength
or orientation.  The stability criterion,
\begin{equation}\label{stab}
({\bf k\cdot v_A})^2 > -  {d\Omega^2\over d\ln R},
\end{equation}
includes the magnetic field through the Alf\'ven speed ${\bf v_A}$ but
only in combination with a wavenumber vector ${\bf k}$. Hence, for any
magnetic field, however weak, one can find an unstable wavenumber so
long as $\Omega$ decreases with $R$.

The stability criterion for a Keplerian disk is 
\begin{equation}
{2\pi\over\lambda_{MRI}} v_A > {\sqrt 3} \Omega ,
\end{equation}
which indicates that the unstable wavelengths, $\lambda_{MRI}$,
are those that are
longer than the distance an Alfv\'en wave travels in one orbit.
The growth rate of the MRI is proportional to $k \cdot v_A$, peaking
at a value of ${3\over 4} \Omega$ for Keplerian disks for a wavelength
\citep{BH98}
\begin{equation}
\lambda_{MRI} = {4\over\sqrt 15} {2\pi v_A\over\Omega}.
\end{equation} 

As a practical matter, for an increasingly weak field, $\lambda_{MRI}$
should eventually fall to a scale small enough that ideal MHD no
longer applies.  When the MRI wavelengths are comparable to the
resistive scale, for example, the field diffuses through the gas
faster than the instability can grow, resulting in stabilization.
At the other extreme, namely the strong field limit, the shortest
unstable wavelength must fit into the disk.  When $\lambda_{MRI}$
is comparable to the scale height of the disk, $H$, then $v_A \sim
c_s$, where $c_s$ is the sound speed.  This is a strong field indeed,
and this condition is often expressed in terms of the plasma $\beta$
parameter, which is the ratio of the thermal to magnetic pressure,
as $\beta \le 1$.   Strong fields do not necessarily mean a stable,
quiet disk, however.  First, when $\beta < 1$ the fields are
dynamically important and can transport angular momentum directly,
possibly through a wind or a jet \citep[see, for example,][]{Lesur13}.  
Second, the stability condition
on a toroidal field requires the Alfv\'en speed to be comparable
to the {\it orbital} speed, rather than the sound speed; confining
fields of this strength within a disk would seem to be problematic.

To conclude, disks for which the gas has sufficient conductivity
as to be magnetized will be linearly unstable to the MRI.  The
action of the MRI is to transfer angular momentum outward through
the disk, precisely what is needed for accretion.

\subsection{MRI-driven turbulence}
\label{sec:3}

The linear analysis establishes that Keplerian orbits are unstable
in the presence of a weak magnetic field.  For any further understanding
of the properties of an unstable disk we must turn to numerical
simulations.  Accretion simulations can be global or local.
Global simulations model the whole disk, or at least a region of
significant radial extent.  Local simulations, on the other hand,
attempt to focus on the properties of the gas within a small region
of the disk.  For accretion disks, local simulations use an approximation
known as the {\it shearing box} \citep{HGB95}.  The shearing box
domain is assumed to be centered at some radial location $R$ that
has an orbital frequency $\Omega(R)$.  The size of the box is taken
to be much less than $R$ so that one can assume a local Cartesian
geometry.  The tidal gravitational and Coriolis forces are retained.
Radial boundary conditions are established by assuming that the box
is surrounded by identical boxes sliding past at the appropriate
shear rate.  Through use of the shearing box approximation one can
study the details of MRI-driven turbulence on scales that are much
smaller than the scale height of the disk or the radial distance
from the central star.  

Extensive shearing box simulations have led to the following general
conclusions: 

\begin{itemize} 

\item[$\bullet$] The MRI leads to MHD turbulence characterized by an anisotropic
stress tensor:  the $R\phi$
magnetic stress component, $T_{R\phi}$, is large, which leads to significant radial transport of
net angular momentum \citep{HGB95}.

\item[$\bullet$]  A net Reynolds stress is also present in the turbulence, and
it too transports angular momentum, but the Maxwell stress is
consistently larger than the Reynolds stress by a factor of 3--4.
If the magnetic fields are removed, the turbulence quickly dies
out \citep{HGB95}.

\item[$\bullet$] Time and space variations in the turbulence can be large.
The turbulence is chaotic \citep{Winters03}.  Characterizing the
stress in terms of an $\alpha$ parameter makes sense only in a space- and time-averaged sense.

\item[$\bullet$] The stress is proportional to the magnetic
pressure, $T_{R\phi} = \alpha_{mag} P_{mag}$, with an $\alpha_{mag}
= 0.4$--0.5. This implies that the traditional Shakura-Sunyaev value
is $\alpha \sim 1/\beta$, where $\beta$ is determined by $P_{mag}$,
the total magnetic pressure
in the turbulent flow.

\item[$\bullet$] The magnetic energy typically saturates at a value
$\beta \sim 10$--$100$.  This is somewhat dependent on the character
of the background magnetic field.   Simulations with net vertical
field tend to saturate at stronger field values.  

\item[$\bullet$] The local model is valid in so far as the energy released by
the stresses is thermalized promptly, on an eddy turnover time-scale
of order $\Omega^{-1}$ \citep{Simon09}.  The rapid thermalization
of the energy is required for $\alpha$-disk theory to be applicable
\citep{BalPap99}.

\item[$\bullet$] The MRI acts like an MHD dynamo in that it sustains
positive magnetic field energies in the face of considerable
dissipation \citep{HGB96}.  Further, stratified shearing box simulations
exhibit a behavior that can be modeled as an $\alpha$--$\Omega$
dynamo \citep{Brandenburg95,Gressel10,GG11}.  This is, however, a small-scale 
dynamo; shearing box simulations have not provided any evidence for the 
existence of a dynamo capable of producing a large scale field.

\item[$\bullet$] In simulations that include resistivity and viscosity, the
amplitude of the MRI-driven turbulent fluctuations and the resulting angular momentum
transport are a function of magnetic Prandtl number.  Tubulence is not sustained
for $P_{\rm M} < 1$ \citep{Lesur07,Simon09b}, i.e., when the resistivity is greater than the viscosity.  

\end{itemize}

The local simulations tell us a considerable amount about the MRI
and the resulting turbulence, but they cannot address questions of
a global nature, such as the net accretion rate into the central
star, losses due to winds and outflows, and the mechanisms by which
jets might be generated. We know, however, that disks will be
magnetized, and this implies the possibility of jet generation from
the disk. In the next section we discuss the modeling of such jets.

%====================================================================================
\section{Magnetic Jets from Disks - Theory and Simulations}
\label{sec:jets}

For jet theory and simulation researchers typically consider only
a subset of the issues listed in Section \ref{sec:intro} and employ
a variety of simplifications. An early example of a jet simulation
that focused on the propagation of the jet and interaction with the
surrounding medium is given by \cite{NWSS82} who carried out some
of the first high-resolution simulations of hydrodynamic axisymmetric
cylindrical jet flows. Some of the earliest simulations of questions
related to jet launching, collimation in relation to disk properties
were given by \cite{US84} \citep[see also][]{SU85,US85}.

Since those early efforts, considerable work has been done on MHD
jet modeling.  One may distinguish i) between steady-state models
and time-dependent numerical simulations, and also ii) between
simulations considering the jet formation only from a fixed-in-time
disk surface and simulations considering also the launching process,
thus taking into account disk and jet evolution together.

Clearly, many jet properties depend on the mass loading of the jet,
which can only be inferred from a treatment of the accretion-ejection 
process. 
While numerical simulations of the accretion-ejection structure potentially 
provide the time-evolution of the launching process, a number of constraints 
that were found, had been discovered already by previous steady state 
models (see below).

\subsection{Magneto-centrifugal Disk Winds}

Steady-state modeling of magneto-centrifugally launched disk winds
have mostly followed the self-similar
\cite{BP82} approach (e.g. \citealt{1994ApJ...429..139C, 1994ApJ...432..508C}). 
Some fully two-dimensional models have been proposed
\citep{1992ApJ...394..117P, 1993ApJ...415..118L}, including
some that take into account the central stellar dipole
\citep{1995A&A...300..791F}.  
Further, some numerical solutions
have been proposed by e.g., \cite{1993ApJ...410..218W, 2010MNRAS.401..479K,
2011MNRAS.412.1162S} in weakly ionized accretion disks that are
threaded by a large-scale magnetic field as a wind-driving accretion
disk.  
They have studied the effects of different regimes for
ambipolar diffusion or Hall and Ohm diffusivity dominance in these
disk.  
Self-similar steady-state models have also been applied
to the jet launching domain \citep{1993ApJ...410..218W,
1993A&A...276..625F, 1995ApJ...444..848L, 1997A&A...319..340F,
2000A&A...361.1178C} connecting the collimating outflow with the
accretion disk structure.
In particular, the fact that large scale magnetic fields need to be 
close to equipartition in order to launch jets via the Blandford-Payne
mechanism is now well accepted, but was first established 
with self-similar studies 
\citep{1993A&A...276..625F, 1995ApJ...444..848L, 1997A&A...319..340F}.
A similar comment can be made also concerning the turbulent magnetic diffusivity 
required in accretion disks.

In addition to the steady-state approach, the magneto-centrifugal jet
formation mechanism has been the subject of a number of time-dependent
numerical studies.  In particular, \citet{1995ApJ...439L..39U} and
\citet{1997ApJ...482..712O} demonstrated the feasibility of the MHD
self-collimation property of jets.  Among these works, some studies
have investigated artificial collimation \citep{1999ApJ...516..221U}, a
more consistent disk boundary condition
\citep{1999ApJ...526..631K,2005ApJ...630..945A}, the effect of
magnetic diffusivity on collimation \citep{2002A&A...395.1045F}, the
impact of the disk magnetization profile on collimation
\citep{2006ApJ...651..272F, 2006MNRAS.365.1131P}, or the impact of
reconnection flares on the mass flux in jets from a two-component
magnetic field consisting of a stellar dipole superposed on a disk
magnetic field \citep{2009ApJ...692..346F}.  

In the context of core-collapse gamma-ray bursts, \citet{Tchekhovskoy-08}
studied the acceleration of magnetically-dominated jets confined
by an external medium and demonstrated that jets gradually accelerate
under the action of magnetic forces to Lorentz factors $\Gamma
\gtrsim 1000$ as they travel from the compact object to the stellar
surface. However, \citet{Kom-09} pointed out an important problem:
such confined, magnetized jets were too tightly collimated for their
Lorentz factors to be consistent with observations. Does this mean
that gamma-ray burst jets are not magnetically powered? It turns
out that magnetized jets are still viable: as they exit the star
and become deconfined, they experience an additional, substantial
burst of acceleration that brings their properties into agreement
with observations \citep{2010NewA...15..749T}.

In the context of stellar jets, \citet{2011ApJ...728L..11R} studied
the large-scale jet formation process spanning the whole range from
the disk surface out to scales of more than 1000 AU.  Further
extensions of this model approach have included the implementation
of radiation pressure by line forces as applied to jets in AGN
\citep{2000ApJ...543..686P, 2004ApJ...616..688P} or massive young stars
\citep{2011ApJ...742...56V}, which may well affect acceleration and
collimation of the jet material.  Simulations of relativistic MHD
jet formation were presented by \cite{2010ApJ...709.1100P} finding
collimated jets just as for the non-relativistic case (above).
Applying relativistic polarized synchrotron radiative transport to
these MHD simulation data through postprocessing yields mock
observations of small-scale AGN jets \citep{2011ApJ...737...42P}.

The stability of the jet formation site has been studied using 3D
simulations for the non-relativistic \citep{2003ApJ...582..292O}
and the relativistic case \citep{2013MNRAS.429.2482P}.  Self-stabilization
of the jet formation mechanism seems to be enforced by the magnetic
``backbone'' of the jet, the very inner highly magnetized axial jet
region.  These studies complement simulations investigating the
stability of jet propagation on the asymptotic scales.  Here, usually
a collimated jet is injected into an ambient gas distribution,
either for the non-relativistic
\citep{1992ApJ...389..297S,1993ApJ...403..164T, 1994ApJ...433..746S,
2000ApJ...540..192S} or the relativistic \citep{2010MNRAS.402....7M,
2008A&A...486..663K} case.

In the aforementioned studies, the jet-launching accretion disk is
taken into account as a boundary condition, {\em prescribing} a
certain mass flux or magnetic flux profile in the outflow.  This
may be a reasonable setup in order to investigate jet formation,
i.e. the acceleration and collimation process of a jet. However,
such simulations cannot tell the efficiency of mass loading or
angular momentum loss from disk to jet, or cannot determine
which kind of disks launch jets and under which circumstances.

It is therefore essential to extend the jet formation setup and
include the launching process in the simulations --- that is, to
simulate the accretion-ejection transition.  Clearly, this approach
is computationally much more expensive.  The typical time scales
for the jet and disk region differ substantially;  disk physics
operates on the Keplerian time scale (which increases as $R^{3/2}$),
and on the even-longer viscous and the diffusive time scales. The
jet follows a much faster dynamical time scale.  As this approach
is also limited by spatial and time resolution, jet launching simulations to
date have employed a rather simple disk model
---namely, an $\alpha$-prescription for the disk turbulent 
magnetic diffusivity and viscosity, and without considering radiative 
effects.

Numerical simulations of the launching of MHD jets from accretion disks
have been presented by \citet{1998ApJ...508..186K,2002ApJ...565.1035K}
and \citet{2002ApJ...581..988C, 2004ApJ...601...90C}.  These
simulations treat the ejection of a collimated outflow out of an
evolving disk through which a magnetic field is threaded.  To prevent
the field from accreting itself, the disk is resistive, allowing
the field to slip through.  \citet{2007A&A...469..811Z} further
developed this approach with emphasis on how resistive effects modify the
dynamical evolution.  An additional central stellar wind was
considered by \citet{2006A&A...460....1M}.  Further studies have
concerned the effects of the absolute field strength or the
field geometry, in particular investigating field strengths around
and below equipartition \citep{2005ApJ...621..921K, 2009MNRAS.400..820T,
2010A&A...512A..82M}.  These latter simulations follow several
hundreds of (inner) disk orbital periods, providing sufficient time
evolution to also reach a (quasi) steady state for the fast jet
flow.

Parameter studies have been able to disentangle the effects of
magnetic field strength (magnetization) and magnetic diffusivity
(strength, scale height) on mass loading and jet speed
\citep{2012ApJ...757...65S}. 
\citet{2013MNRAS.428.3151T} considered how entropy affects the 
launching process, in particular how disk heating and cooling 
influence the launching process.  They find that heating at the disk surface enhances the
mass load, as predicted in the steady state modeling by \citet{2000A&A...361.1178C}.

Most jets and outflows are observed as {\em bipolar} streams. Very
often, observations show asymmetrical jets
and counterjets. For protostellar jets one exception is HH\,212, which
shows an almost perfectly symmetric bipolar structure
\citep{1998Natur.394..862Z}.  For relativistic jets Doppler beaming
may play a role for the observed jet asymmetries, and, in fact, many
well-behaved jets can be modeled on the assumption that Doppler
beaming dominates the apparent asymmetry \citep{Laing+Bridle14}.
However, environmental asymmetries must be also considered.  It is therefore
interesting to investigate the evolution of both hemispheres of a
{\em global} jet-disk system in order to see whether and how a
global asymmetry in the large-scale outflow can be governed by the
disk evolution.  \citet{2013ApJ...774...12F} have been able to
trigger jet asymmetries by disturbing the hemispheric symmetry of
the jet-launching accretion disk and find mass flux or jet velocity
differences between jet and counter jet of up to 20\%.

\citet{2003A&A...398..825V} and \citet{2004A&A...420...17V}
investigated the origin of the magnetic field driving the jet
by including a mean-field disk dynamo in the star or in the disk.
Asymmetric ejections of stellar wind components were found from 
offset multi-pole stellar magnetospheres \citep{2010MNRAS.408.2083L}.
The most recent work considers the time ($> 5\times 10^5$ dynamical time
steps) evolution on a large spherical grid ($>2000$ inner disk
radii), including the action of a mean-field disk dynamo that builds
up the jet magnetic field. A variable dynamo action may cause the
time-dependent ejection of jet material \citep{2014..stepanovs..a,
2014..stepanovs..b}.

The main jet launching processes seem to be well understood,
in the sense that a large number of independent simulation studies
give consistent results. Several
issues are not yet resolved, however.
These include:
(i) the origin of jet knots, and (ii) the coupling of the small
scale disk physics to the global jet outflow.  In the end there is
a good chance that the answers to both questions are interrelated.
A full numerical simulation that addresses these questions would
require substantial effort, including full 3D,  high resolution,
and more complete physics, in particular the treatment
of thermal effects. First steps in this direction have
already been taken.

Concerning the origin of jet knots, it is still unclear whether
knots are signatures of internal shocks of jet material launched
episodically with different speed,  external
shocks of jet material with the ambient medium, or re-collimation
shocks.  One observation supporting the intrinsic origin of jet
knots is that of the jet of HH\,212, which shows an almost perfectly symmetric
bipolar structure with an identical knot separation for jet
and counter jet \citep{1998Natur.394..862Z}.  Such a structure can
only be generated by a mechanism intrinsic to the jet source.

The knot separation in protostellar jets typically corresponds to
time scales of $\tau_{\rm kin} \simeq 10-100$ yrs.  
By contrast, a typical time scale for the jet launching area would be 
about 10-20 days, that is the Keplerian period close to an inner disk 
radius of 0.1\,AU.
This time scale increases up to one year if the jet launching 
radius is larger, say up to 3-5 AU \citep{2014prpl.conf..451F}. 
However, outflows launched from such large radii would probably not
achieve the high velocities of $300-500\,{\rm km\ s^{-1}}$ observed for jets.
Thus, the mechanism responsible for the jet knots must be intrinsic to the
disk, and triggered by a physical processes on a rather long time
scale.  Candidates are disk thermal (FU Orionis) or accretion
instabilities, a mid-term variation of the jet-launching magnetic
field, or MRI-active/dead disks.  

\begin{figure}[t!]
\centering
\includegraphics[width=0.8\linewidth,clip=true]{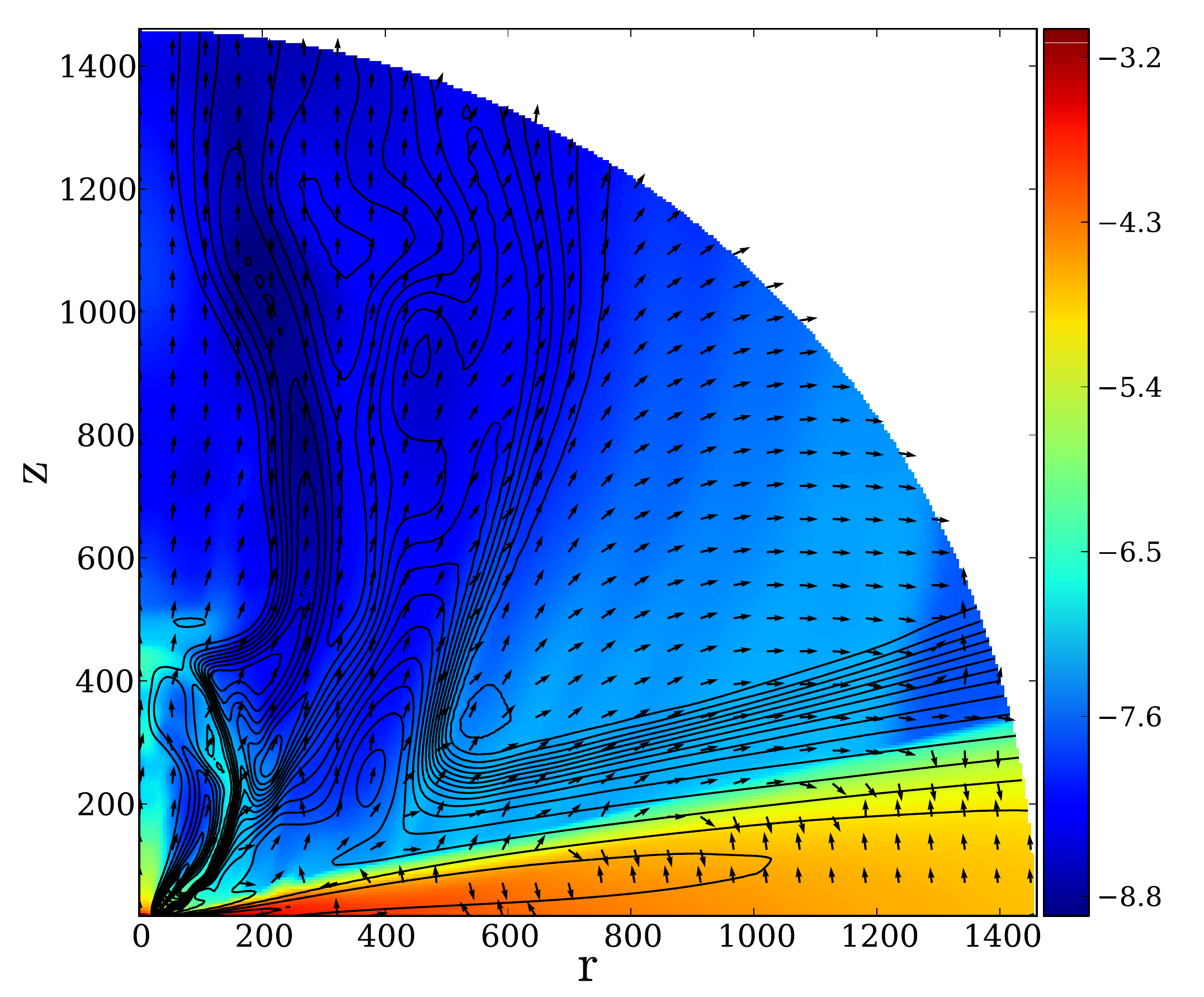}
\includegraphics[width=0.8\linewidth,clip=true]{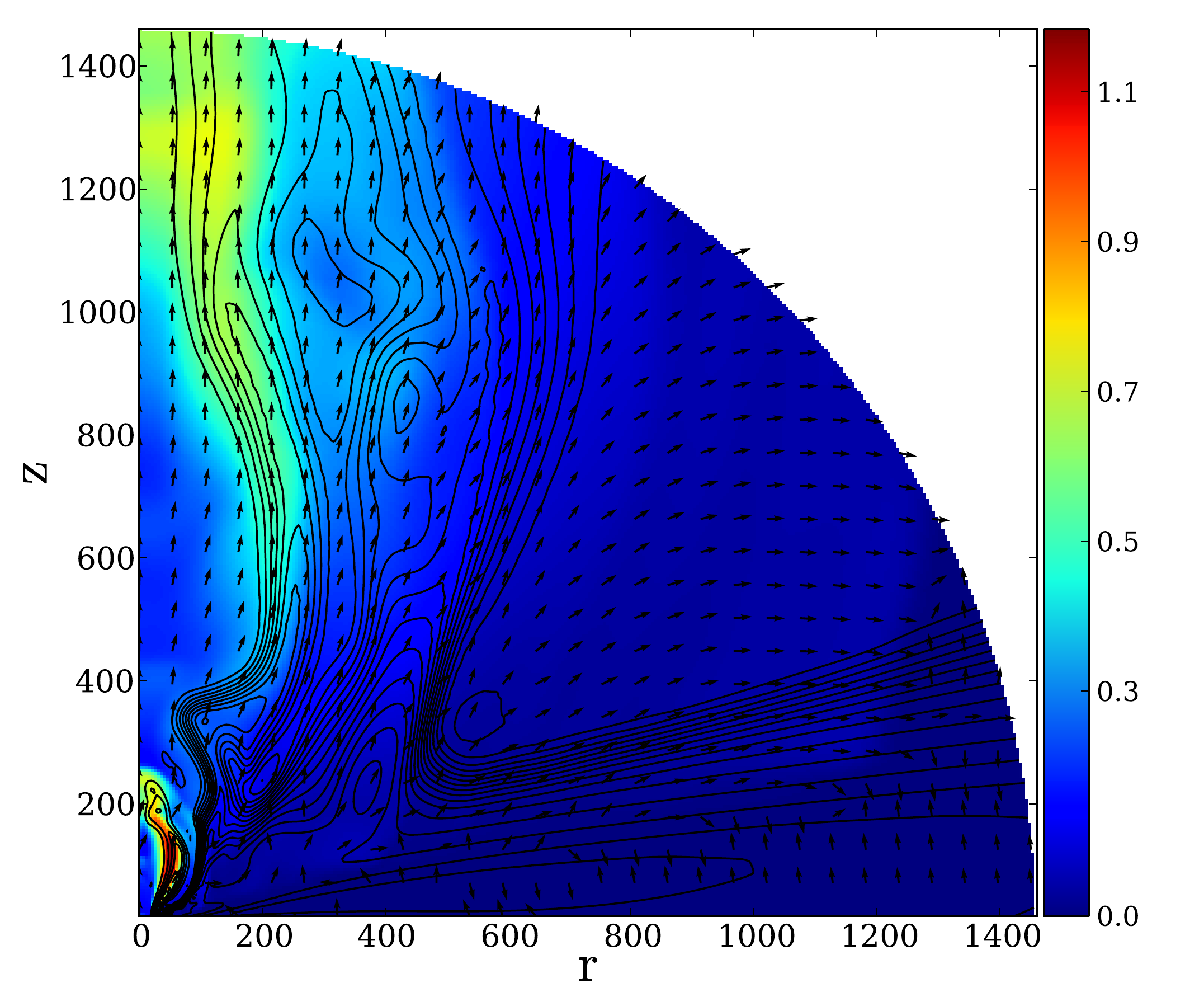}
\caption{Time evolution of the inflow-outflow structure for a
dynamo-generated disk magnetic field.  Shown is the density (top,
colors, in logarithmic scale), and poloidal velocity (bottom, colors,
linear scale), the poloidal magnetic flux (thin black lines), and
the poloidal velocity vectors.  The dynamical time step is about
$10^5$, while the dynamo time scale (switch on/off) is $10^3$.  Two
ejections of material launched at $\Delta t = 2000$ are
more easily seen in the velocity figure (below), than in the top
figure, which shows a slightly later time when the preceding knot has left the
grid already. Length units is the inner disk radius of $\sim 0.1$ AU.
Figure taken from \citet{2014..stepanovs..a}.
}
\label{fig:knots}
\end{figure}

The simulations of \cite{2014..stepanovs..b} provide an example of
episodic knot ejections.  They apply a $\alpha$-$\Omega$ mean-field
disk dynamo to generate the jet launching magnetic field. 
In contrary, all previous simulations of this kind start with a 
strong initial magnetic field distribution.
The episodic ejections are triggered by switching on and off the dynamo
term every 2000 dynamical time steps (Fig.\ref{fig:knots}).

The idea of a connection between the small scale disk physics and
the global disk outflow is motivated by the fact that jet launching
relies on both the existence of a large scale magnetic field
and the existence of a (turbulent) magnetic diffusivity, which allows
for accretion through the field lines, for angular momentum transfer,
and, essentially, also for the mass loading onto the jet magnetic
field.  The disk turbulence is a result of the magnetorotational
instability (MRI, see above) and thus of the small scale physics.

First results combining a mixed numerical-analytical treatment
indicate that large scale MRI modes may produce magnetically driven
outflows, however, these flows seem to be 3D-unstable
\citep{2013A&A...550A..61L}, and thus may be a transient effect.
On the other hand, local numerical simulations suggest that 
magnetocentrifugal winds can be launched when the MRI is suppressed,
in spite of the low magnetization $\sim 10^{-5}$ 
\citep{2013ApJ...767...30B, 2013ApJ...769...76B}. 
In this case, outflows are launched from the disk surface region where 
the plasma $\beta$ is about unity. 
This is also what \citet{2010A&A...512A..82M} have observed in global 
launching simulations.
\citet{2013A&A...550A..61L} have found that the MRI near equipartition 
does not lead to turbulence, but can be responsible for jet launching.

%The topic of the disk-jet connection is
%discussed further in the following section.

The overall goal is still to answer the question: what kind of disks
launch jets and what kind of disks do not?  It is clear from
statistical arguments for both protostars and extragalactic sources
that jets are a relatively short-lived phenomenon.  The typical jet
propagation time scale for protostellar jets is about $\tau_{\rm
dyn} \equiv L_{\rm jet} / V_{\rm jet} \simeq 10^4$ yrs, while
protostellar disks live some $10^6$ yrs.  A similar argument can
be raised for AGN jets based on the fact that there are far more
radio-quiet AGN (quasars) than radio-loud AGN (jet sources), although
at least some types of radio-loud AGN appear to be almost continuously
in an `on' state. So far, modeling has not provided an answer to this
question.

% JH

\section{Black Hole Accretion, Fields and Jets}
\label{sec:accretion-jet}

\subsection{Simulating the black hole jet}

As we have seen, a plethora of models and simulations have
demonstrated the ability of magnetic fields to launch, accelerate and
collimate jet outflows. But simulations designed to study jets have,
as discussed in the previous section, typically used one or another
specific set of initial or boundary conditions that pre-suppose a disk structure, a
mass-energy flux, a magnetic field configuration, or some 
combination of these.  An alternative approach is to focus not on the jet, but
on on the accretion disk itself and see when and under what
circumstances a jet might {\it develop} from the flow.

Most global simulations begin with an orbiting axisymmetric hydrostatic
equilibrium (torus) a few tens of gravitational radii from the black
hole. The initial magnetic field is entirely contained within the
matter, so that there is no net magnetic flux and no magnetic field
on either the outer boundary or the event horizon. A favorite
configuration consists of large concentric dipolar loops
\citep{Hawley:2000,mg04,dVhkh05,hk06,McKinney:2006}.  The development
of the MRI subsequently leads to the establishment of outflows from the disk. 
The wind generation seems to be largely
an outcome of thermodynamics. In simulations that do not include
cooling, the combination of the heat released from accretion and the
buildup of magnetic pressure lead to pressure-driven outflows.  In the 
absence of a large scale vertical field through the disk, however,
this wind is neither collimated nor unbound.

Global simulations have nevertheless produced jets that are consistent
with the Blandford-Znajek mechanism, through the creation of a
substantial axial field that threads the black hole.  For example, in
a model that begins with a torus containing dipolar field loops, 
differential rotation
rapidly generates toroidal field.  This field increases the total
pressure within
the torus, driving the its inner edge inward. As the field is
stretched radially, the Maxwell stress transports angular momentum
outward, further enhancing the inflow.   Subsequent
evolution carries the inner edge of the disk, and the magnetic field
within,  down to the black hole. The field expands
rapidly into the nearly empty \textit{funnel} region along the black
hole axis.  Gas drains off the field lines and into the hole leaving
behind a low $\beta$ axial field.  Frame dragging by the
rotating hole powers an outgoing Poynting flux. The power of the
Poynting flux jet is determined by the spin of the black hole and
the strength of the magnetic field \citep{BZ-77},
as we discuss below. The simulated jet's power matches well with
the predictions of the Blandford-Znajek model
\citep{mg04,McKinney:2005,Tchekhovskoy-11}. 

The matter content within the Blandford-Znajek jet is very small;
the angular momentum in the disk material is too great to enter the
axial funnel.  In the simulations the jet density is typically set by the value assigned to the numerical vacuum.  How these jets become 
mass-loaded in Nature is still something of an open issue.
 Since the boost factor of the jet is determined by
the mass loading, $\Gamma$ is typically large, e.g., $\Gamma \sim
10$ \citep{McKinney:2005}. No particular weight can be given to any
value of $\Gamma$ found in simulations, although it is safe to say
that the feasibility of values as large as those observed (Section
\ref{sec:obs}) has been demonstrated.

Although there is no appreciable matter within the funnel, some of
the field lines that lie just outside the funnel pass through the
accretion flow, and those field lines can accelerate matter outward
in a collimating flow.  This ``funnel wall jet'' was observed in
early pseudo-Newtonian simulations \citep{Hawley:2002}, and is a
regular feature of fully relativistic jet simulations, e.g.,
\cite{hk06}.

Because the field lines in the jet rotate, the jet carries angular
momentum as well as energy away from the black hole.  The angular
momentum flux can be large, comparable to that brought down by
accretion \citep{gsm04,hk06,bhk08}.  One potentially important
consequence of this was noted by \cite{gsm04} who pointed out that
electromagnetic angular momentum flux in the jet rises so steeply
with black hole spin that it may limit $a/M$ to $\simeq 0.93$
\citep[see also][]{mg04,hk06,bhk08} or even down to $\lesssim0.1$, as
we discuss below \citep{Tchekhovskoy-2012a,Tchekhovskoy-15}.  The
possibility that magnetic fields could play such an important role in
black hole physics is certainly in keeping with the spirit of this
workshop!

For a magnetic field strength $B$ at the event horizon, jet power
is (roughly) given by the magnetic energy density, $B^2$, times jet
cross-section at its base, $r_g^2$ (where $r_{\rm g} = GM/c^2$ is
the black hole gravitational radius and $M$ its mass), times the
speed $v\sim c$ at which the field moves out,
\begin{equation}
P_{\rm jet} \sim (a/M)^2 B^2 r_g^2 c \sim  (a^2 c/M^2r_g^2)\Phi^2 = k\Phi^2,
\label{eq:pbz}
\end{equation}
where the effect of black hole rotation is in the $(a/M)^2$
pre-factor, and the magnetic field is expressed in terms of the 
flux $\Phi \sim B r_g^2$.  Since in the course of an observation $M$
and $a$ usually do not change appreciably, the factor $k = a^2
c/M^2r_g^2$ is a constant, and all changes in jet power
are due to changes in the magnetic flux,~$\Phi$.
 
\begin{figure}[t!]
\includegraphics[width=\linewidth]{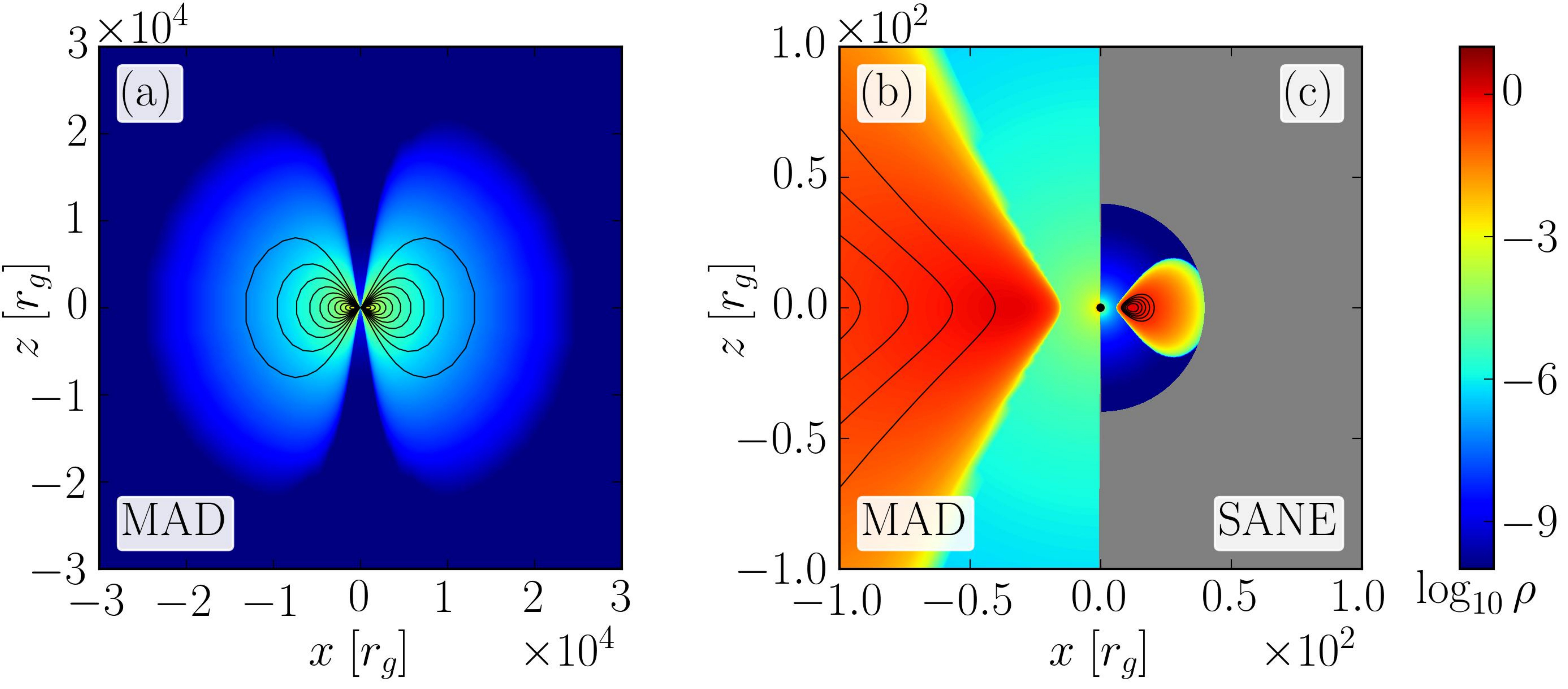}
\caption{ Panels (a) and (b)
  show a meridional slice of density with color (red shows high,
  blue low density, see color bar) and magnetic field lines with black
  lines for the initial conditions of the simulations of \citet{Tchekhovskoy-11}
  The initial magnetic field is fully contained within the
  torus, and have a sufficient amount of
  magnetic flux to saturate the black hole with magnetic flux and to
  lead to the development of a magnetically-arrested disk (MAD). 
  Panel (c) contrasts this initial condition with a typical initial condition in 
  earlier simulations where
  the torus extent is smaller, and the magnetic
  flux is reduced.
  This is labelled ``SANE'' ICs, for ``standard and normal evolution''
  \citep{2012MNRAS.426.3241N}. Figure adapted from
  \citet{Tchekhovskoy-15}.}
\label{fig:madvssaneics}
\end{figure}

What is the possible range of $\Phi$ and the corresponding range
of $P_{\rm jet}$?  The lower limit is clear: zero magnetic flux
leads to zero power.  But what sets the maximum value of $\Phi$?
In the simulations of \cite{bhk09} the black hole field strength
is comparable to the (mostly gas) pressure in the inner disk,
$\beta \sim 1$, suggesting that, in general, equipartition holds.  
This further suggests that
even higher black hole magnetic field strengths are possible for more
energetic accretion flows.
\citet*{Tchekhovskoy-11} carried out numerical simulations of black
hole disk-jet systems designed specifically to address the question
of maximum field strength.  The simulation begins with an initial
thick, hot, large scale torus (as shown in Fig.~\ref{fig:madvssaneics}a,b).
Because of the large torus size, it contained a particularly large
amount of magnetic flux, much larger than in previous work.
Fig.~\ref{fig:madvssaneics}a,b shows the initial condition, compared
with initial conditions typically used in previous simulations
(Fig.~\ref{fig:madvssaneics}c).  Using these initial conditions
\citet{Tchekhovskoy-11} found that magnetic flux on the black hole
saturated at an equipartition point, one where the magnetic
pressure was sufficient to halt accretion against the inward pull
of gravity. In other words, the dynamically-important magnetic flux
obstructed the accretion and led to what is dubbed a \emph{magnetically
arrested disk}, or a ``MAD''; in such a condition no further flux
could be carried down to the hole.  In contrast, models where
magnetic field falls short of saturating the black hole were referred
to as ``SANE'' accretion models, for ``standard and normal evolution''
by \cite{2012MNRAS.426.3241N}.
In the MAD simulation $\Phi$ and $P_{\rm jet}$ are greatly increased, as we
discuss below
\citep{Tchekhovskoy-11,Tchekhovskoy-2012a,2012MNRAS.423.3083M,Tchekhovskoy-15}.

How do such strong magnetic fields manage to stay in force-balance
with the disk?  As in previous disk simulations, e.g., \cite{bhk09},
the black hole field strength is comparable to the total
pressure in the disk.  In MADs, magnetic and gas pressures in the
disk are comparable, $\beta\sim1$,  but the much stronger
black hole magnetic field in the MAD case compresses the inner disk
substantially, so the inner regions of MAD disks are actually
over-pressured compared to previous simulations by a factor of $\sim 5{-}10$.

\begin{figure}
\begin{center}
\includegraphics[height=0.65\textwidth]{diskjet3d.pdf}\hfill
\includegraphics[height=0.65\textwidth]{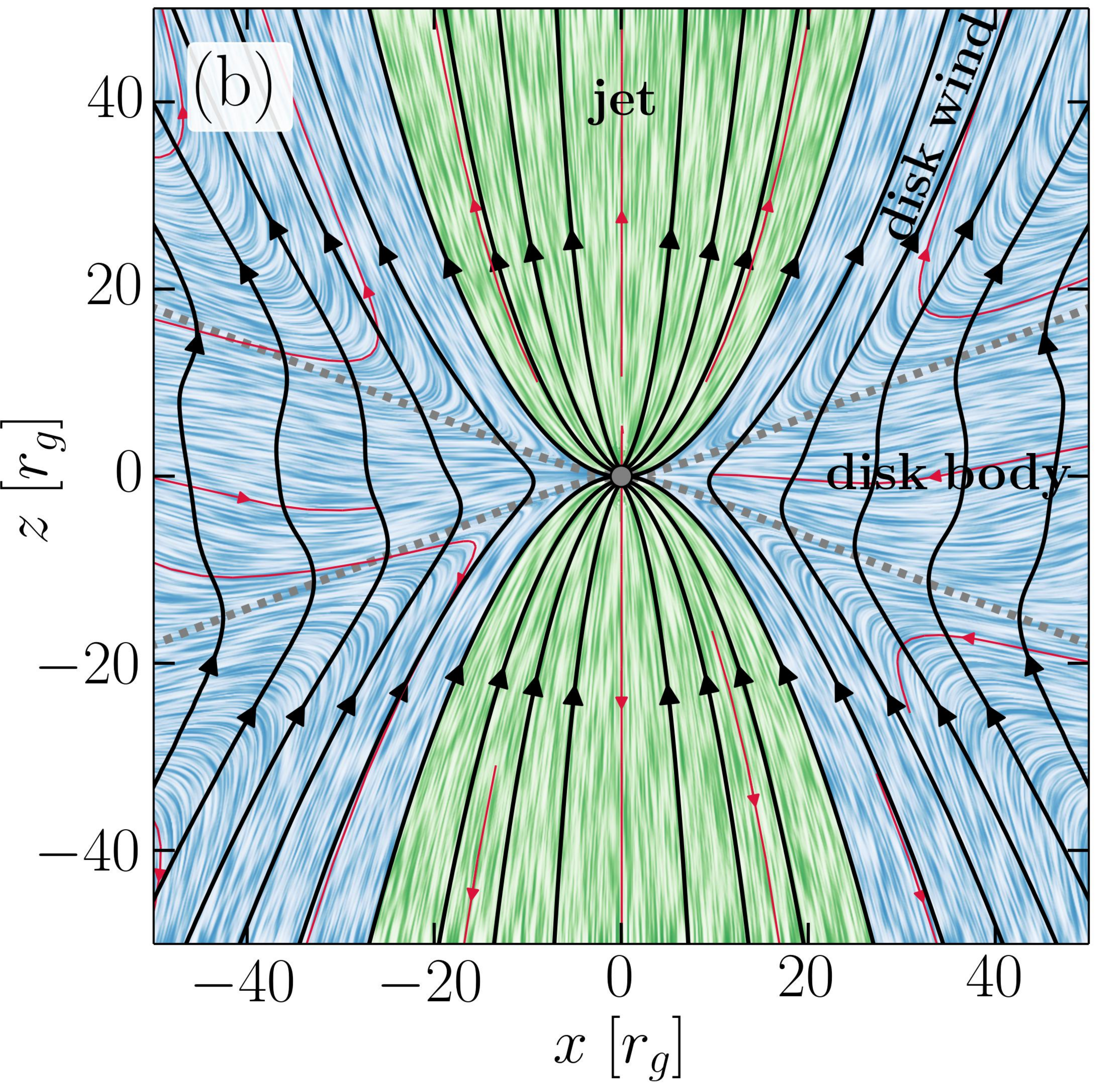}
\end{center}
%\vspace{-0.5cm}
\caption{\label{fig:dja} {\bf [Panel (a)]:} A 3D rendering of a MAD
  disk-jet simulation. Dynamically-important magnetic fields are twisted by
  the rotation of a black hole (too small to be seen in the image) at the
  center of an accretion disk.  The azimuthal magnetic field component
  clearly dominates the jet structure. Density is shown with color:
  disk body is shown with yellow and jets with cyan-blue color; 
  jet magnetic field lines are cyan bands.  The image size is
  approximately $300r_g\times 800r_g$. {\bf [Panel (b)]:} Vertical
  slice through a MAD disk-jet simulation averaged in time and
  azimuth. Ordered, dynamically-important magnetic fields remove the
  angular momentum from the accreting gas even as they obstruct its
  infall onto a rapidly spinning black hole ($a/M=0.99$). Gray filled circle
  shows the black hole, black solid lines show poloidal magnetic field lines,
  and gray dashed lines indicate density scale height of the accretion
  flow, which becomes strongly compressed vertically by black hole
  magnetic field near the event horizon.  The symmetry of the
  time-average magnetic flux surfaces is broken, due to long-term
  fluctuations in the accretion flow. 
  Streamlines of velocity are depicted both as 
  thin red lines and with colored ``iron filings''. which are better
  at indicating the fine details of the flow structure.  The flow
  pattern is a standard hourglass shape: equatorial \emph{disk inflow}
  at low latitudes, which turns around and forms a \emph{disk wind
  outflow} (labeled as ``disk body'' and ``disk wind'',
  respectively, and highlighted in blue), and twin polar \emph{jets}
  at high latitudes (labeled as ``jet'' and highlighted in
  green). Figure taken from \citet{Tchekhovskoy-15}.  }
\end{figure}

Figure~\ref{fig:dja} shows the structure of the simulated disk-jet
system.  An accretion disk around the central black hole leads to
a pair of jets that extend out to much larger distances than the
black hole horizon radius and are collimated into small opening
angles by outflows from the disk (not shown). Jet magnetic field
is predominantly toroidal, which reflects the fact that the jets
are produced by the rotation of the black hole, which twists the
magnetic field lines into tightly-wound helices.

Even though the accretion disk and the jets turn out to be highly
time-variable, in a time-average sense the structure of the flow
is remarkably simple. The poloidal magnetic field lines are shown
with black solid lines in Fig.~\ref{fig:dja}(b).  The group of field
lines highlighted in green connects to the black hole and makes up
the twin polar jets.  As discussed above, these field lines have
little to no gas attached to them; disk gas cannot cross magnetic
field lines and has too much angular momentum to enter the funnel.
The large Poynting flux and very small inertia yields highly
relativistic velocities.  The field lines highlighted in blue connect
to the disk body and make up the magnetic field bundle that produces
the slow, heavy disk wind, whose power is much smaller than $P_{\rm
jet}$ for rapidly spinning black holes.

\begin{figure}[tH!] \begin{center}
\includegraphics[width=0.65\textwidth]{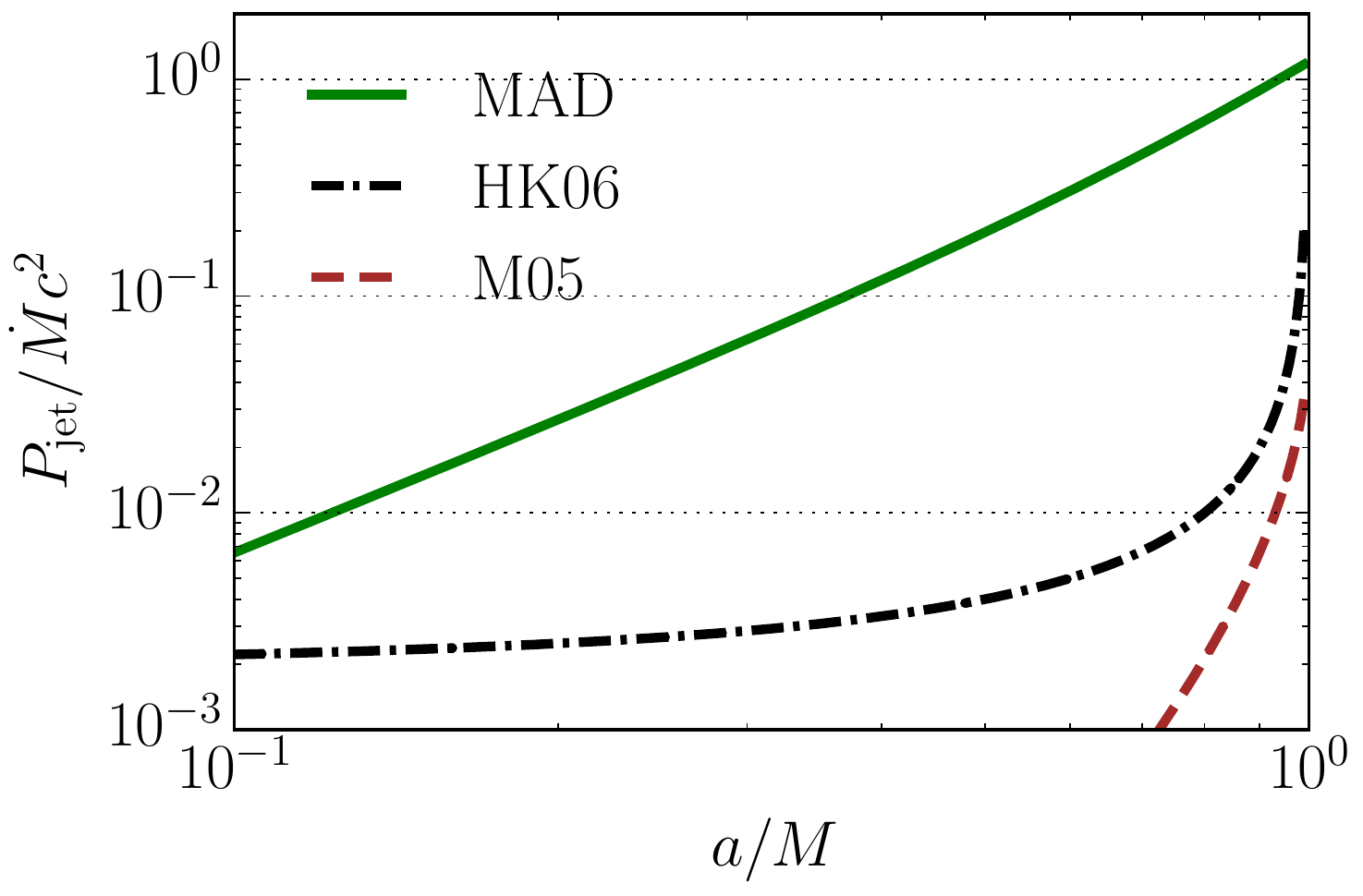}
\end{center} \vspace{-0.5cm} \caption{\label{fig:etajet_comparison}
  Comparison of jet energy 
  efficiency obtained in MAD simulations, $P_{\rm jet}/\dot M c^2$, which
  is shown with green solid line, with previously reported
  approximations of simulated jet power: \cite{hk06}, which is shown with
  black dash-dotted line (HK06), and \cite{McKinney:2005}, 
  which is shown with brown dashed line (M05), plotted over the range
  $0\le a/M \le 0.99$. MADs can produce much more powerful
  jets because they have the maximum possible amount of magnetic
  flux threading the black hole. }
\end{figure}

The jet power can be characterized by the \textit{outflow energy efficiency}
$\eta$ as
\begin{equation}
  \label{eq:efficiency}
  \eta = {P_{\rm outflow} \over \dot Mc^2}.
\end{equation}
For the MAD simulation, the accretion rate $\dot M$ fluctuates
considerably, but does maintain a well-defined average value.  In the
MAD flow, for a rapidly spinning black hole, $a/M = 0.99$, the total
outflow power, $P_{\rm outflow}$, which is the sum of the black hole 
powered jet, $P_{\rm jet}$ and the disk powered wind, exceeds
the accretion power, $\eta > 1$, i.e., \emph{net} energy is extracted
from the black hole \citep{Tchekhovskoy-11}.
This is a promising result since high values of jet power
are consistent with or are preferred by observations
\citep{Rawlings+Saunders91,2010MNRAS.402..497G,2011MNRAS.411.1909F,2011ApJ...727...39M,2014arXiv1406.7420N,2014Nat..515...376}.

The MAD flow hearkens back to the ion-torus model of \cite{Rees:1982}
where the role of the thick torus is to confine and anchor the
fields that extract rotational energy from the hole, and accretion
{\it per se} is near zero. This model was put originally forward
as an explanation for systems where high jet power is seen, but
very little luminosity emerges from the central engine.  In the MAD
flow, accretion continues at a non-zero rate regulated by the 
magnetic interchange instability in the disk and the strength of the
central magnetic field.

Figure~\ref{fig:etajet_comparison} compares an analytic approximation
for $P_{\rm jet}$ in MAD simulations to some earlier simulations
\citep{McKinney:2005,hk06}. Jet efficiency in the MAD case is
$\propto a^2$, suggesting the ratio of magnetic flux squared to mass accretion
rate is roughly constant 
\citep{Tchekhovskoy-2012a,Tchekhovskoy-15}.
If MADs achieve the maximum possible amount of
magnetic flux threading the black hole then they have achieved the
maximum possible jet power for a given mass accretion rate (see
Fig.~\ref{fig:etajet_comparison}).  The other models show a rapid rise
as $a/M$ approaches 1. \cite{hk06} attribute this to greater
local field amplification for the extreme Kerr holes; the field has
``room to grow,'' and is a factor of a few smaller than in the MAD
simulations.

Efficiencies $\eta > 1$  can lead to spindown
of the black hole. For example, in an AGN accreting at $10$\%
Eddington luminosity, the central black hole is spun down to near-zero
spin, $a/M\lesssim0.1$, in $\tau\simeq3\times10^8$~years
\citep{Tchekhovskoy-2012a,Tchekhovskoy-15}. This value is interesting
astrophysically because it is comparable to a characteristic quasar
lifetime (Section \ref{sec:obs}). Over this time period, jets can
extract a substantial fraction of the central black hole spin energy
and deposit it into the ambient medium. The central galaxy in the
cluster MS0735.6+7421 may be one such example \citep{2009ApJ...698..594M}.

\subsection{Origin of the jet field}
\label{sec:advection}

Jets appear to require some large-scale magnetic field, in particular
a net vertical field connected either to the disk itself or the
central star or black hole.  This raises the question of the origin
of that field.  In the jet-forming global simulations, the initial
condition consists of dipole loops embedded within initial orbiting
gas, typically a torus of some thickness.  In essence, the subsequent
evolution (which is a consequence of the initial conditions) transfers
the dipole field down to the black hole.  Other initial field
configurations, such as quadrupolar, toroidal, smaller scale loops,
have been found to be far less effective in creating jets \citep{bhk08}.

Simulations that produce jets rely on favorable initial conditions.
What happens in Nature?  At least two possibilities exist.  One is
that the required field is generated {\it in situ} by the disk
itself through a dynamo process, as a natural consequence of
accretion.  Some example dynamo models were discussed above (e.g.,
Fig.~\ref{fig:knots}).  The other possibility is that the required
field is carried into the central disk/star, from the companion
star in a binary system and from the interstellar medium in an AGN.
The advantage of advection is that a relatively weak net field at
large radius could be significantly amplified simply by the geometric
factor in going from large to small radius.  For example, magnetic fields at the
edge of the sphere of influence of a supermassive black hole, i.e.,
$r\sim100$~pc, are plausibly $B\sim\mu$G. Suppose these fields
maintain their coherence over roughly a similar length scale.  The
magnetic flux contained in patch of field this size, $\Phi_{\rm
patch} \approx 10^{35}$ G cm$^2$, is much larger than the flux
necessary to saturate a black hole with magnetic field, $\Phi_{\rm
MAD} \approx 10^{33.5}$ G cm$^2$ (for a black hole of mass $M =
10^9M_\odot$ accreting at $10$\% of Eddington luminosity
\citealt{nia03,Tchekhovskoy-15}).

Under what general circumstances can a net field be advected in by the
accretion flow?  The question is traditionally framed in the terms used by
\cite{vanB89}, namely that accretion rate is set by the viscosity
$\nu$ and the magnetic diffusion by the resistivity $\eta$.  Whether
or not a field can be advected inward depends on the magnetic Prandtl
number $P_m = \nu/\eta$.  If $P_m$ exceeds 1 then field advection
will not occur \citep[see also][]{1994MNRAS.267..235L}.   In accretion
disks the viscosity and resistivity are not the physical values,
but instead are effective values resulting from the turbulence
\citep{GG09,2009A&A...504..309L,2009A&A...507...19F}.

In global simulations, flux in the initial torus (e.g.
Fig.~\ref{fig:etajet_comparison}) finds its way onto the black hole,
despite starting from a large radius.  This suggests that
geometrically-thick disks are capable of transporting large-scale
magnetic flux inward, at least from the distances of $r\sim100r_g$.
In standard $\alpha$-disk theory, infall rates are proportional to
$(h/r)^2$.  If $h/r$ is close to unity the infall time becomes
relatively rapid, and the inflow rate can exceed the diffusion rate.
Thick, radiatively inefficient flows, then, seem to have the best
chance of transporting flux inward.

The question of whether a turbulent, radiatively efficient disk
could transport flux inward remains an interesting one.  \cite{bhk09}
studied the question through a global simulation of an initial
orbiting torus with an $h/r \sim 0.1$ embedded in a vertical magnetic
field.  They found that there was no net flux transported through
the disk itself; the turbulence transported angular momentum and
mass, but not net flux.  Net flux was nevertheless brought down to
the black hole by a coronal inflow surrounding the accretion disk.
Initially the field lies along cylinders of constant rotation, but
at high altitude $B^2/\rho$ is much larger than deep inside the
disk, and the field is subject to a finite-amplitude version of the
MRI.  The sign relation between the radial field component and the
toroidal field component is the usual one, so the Maxwell stress
has the right sign to transport angular momentum outward.  The
angular momentum flux due to the Maxwell stress is large compared
to the fluid angular momentum density.  Fluid elements at high
altitude, both above and below the equator,  are therefore driven
inward quickly, carrying half-loops of field with them.  When one
loop approaches the central black hole from above and outward while
another approaches the disk from below and outward, their local
field directions are opposed, and they can reconnect.  Reconnection
changes the field topology, creating a closed loop at larger radius
and an open field line at small radius.   The open field line, the
one carrying the net flux, is located where the reconnection occurred,
which is at a radius considerably smaller than the initial radius
of the field line.  Thus, reconnection causes the flux to move
inward in large jumps.  All of this takes place faster than the
mean mass inflow rate within the disk at the vicinity of the flux
line's initial position.

In a similar vein, \cite{Suzuki14} carried out a series of simulations with a relatively thick disk
(sound speed $\sim 0.1$ of the orbital speed) with an initial weak magnetic field.
In these simulations net vertical flux was carried inward by a rapdily 
infalling layer near the disks surface.  All these simluations transport net flux
through a mechanism that is governed by large-scale torques and resulting
rapid inflows, rather than by small-scale turbulence where the turbulent
velocities exceed the mean inward drift velocity associated with
accretion.  It appears plausible, therefore, that net inward flux
transport can occur for certain types of accretion flows.

Testing jet-generation models, including quantitative predictions
for the magnetic flux close to the black hole, fundamentally depends
on observations that can give insights into the field properties both
at the jet base and further out. We discuss those in the following
section.

\section{Observational constraints on jet magnetic fields}
\label{sec:obs}

\subsection{Introduction: basic jet properties}

In this section we ask what the observational constraints are on
the field strengths and configurations in jets, and whether
these observational results can be connected to jet-launching models.
(For a more detailed review of these issues see \cite{Pudritz+12};
parts of that work are summarized in this section.) We focus on
relativistic AGN jets, since these are by far the best-studied of the
systems producing synchrotron radiation, and begin by
setting out some basic properties of the jets in radio-loud AGN.

AGN jets are fast outflows carrying kinetic powers that are estimated to be
between $\sim 10^{42}$ and $\sim 10^{47}$ erg s$^{-1}$
\citep[e.g.,][]{Rawlings+Saunders91}. Bulk speeds are known to be
highly relativistic on small scales ($\Gamma \sim 10$--30, and
perhaps higher in some objects: \citealt{Lister+09}); on kpc scales
some jets have decelerated to sub-relativistic speeds
\citep{Laing+Bridle02} while in more powerful systems jets are at
least mildly relativistic out to hundred-kpc scales
\citep{Mullin+Hardcastle09} and may retain a highly relativistic
spine with $\Gamma \sim 10$ out to those scales, though direct
observational evidence for this remains debatable
\citep[e.g.,][]{Tavecchio+00,Hardcastle06,Marshall+11,Konar+Hardcastle13}.

Jet composition remains an open question. Observations of synchrotron
radiation require relativistic leptons and magnetic fields to be
present, but {\it direct} observational constraints on the presence of
a hadronic component are hard to obtain. Some indirect constraints
suggest that the bulk of the required energy for jets can be carried
by the radiating leptons without any need for additional components on
scales from pc to kpc \citep[e.g.,][]{Wardle+98,Croston+05-2,Wykes+13},
It is certain that all jets must entrain material from stellar
winds in the host galaxy \citep{Bicknell94} so their composition may
change with distance along the jet, and there is some evidence from
pressure balance arguments that this is indeed the case
\citep{Croston+Hardcastle14}.

One important observational constraint on jet launching mechanisms
comes from the fact that they appear to be able to originate in AGN
with a wide range of different properties. The vast majority of
radio-loud AGN are so-called `low-excitation radio galaxies' which
show no evidence at any wavelength for a standard radiatively
efficient accretion disk \citep{Hardcastle+09,Best+Heckman12}.
However, the existence of powerful radio-loud quasars with $L/L_{\rm
  Edd} \sim 1$ in the optical shows that radiatively inefficient
accretion is not directly connected to radio jet activity. In fact
it appears that radio-loud AGN have a very wide range of $L/L_{\rm
  Edd}$, even when jet power is taken into account \citep{Mingo+14}.

Finally, it is important to note that observations of synchrotron
radiation require a particle acceleration mechanism. There is indirect
or direct evidence for local particle acceleration on all scales of
radio-loud AGN up to scales of tens of kpc. Shocks are almost
certainly implicated --- some of the best-understood sites of particle
acceleration, the hotspots of powerful double radio galaxies, are
clearly physically associated with the jet-termination shock and show
properties consistent with simple first-order Fermi acceleration
\citep{Meisenheimer+97}. However, other mechanisms may well be
necessary to explain diffuse particle acceleration distributed on
scales of many kpc \citep[e.g.,][]{Hardcastle+07}. When interpreting
observations, it is necessary to bear in mind that
the particles being observed at a given location may owe their energy
spectrum to acceleration in some region that is spatially quite
distinct.

Because jets can be persistent (ages of the order $10^8$ years have
been estimated for some sources based on spectral or dynamical age
estimates) they produce large-scale, long-lasting structures
consisting of material that has passed up the relativistic jet
\citep{Scheuer74}, generally called lobes or plumes. The physical
conditions in these structures are to some extent determined by the
jet and so they give important clues to aspects of the jets that are
hard to investigate directly. Because the observational techniques
used are rather different, it is usual to discuss the large-scale
lobes, the kpc-scale jets and the pc-scale jets separately, a
convention that we follow in the remainder of this section.

\subsection{Observational tools}\label{subsec:32}

The combination of relativistic leptons and magnetic field gives rise
to synchrotron emission, while relativistic leptons and a photon field
give us inverse-Compton emission. These are the basic tools available
to estimate the magnetic field properties in jets and their products.

Optically thin synchrotron radiation cannot on its own give a
measurement of magnetic field strength. For simplicity, let us
represent the electron energy spectrum as a power law with index $p$,
so that $N(E) = N_0 E^{-p}$. Then the synchrotron emissivity in the
optically thin regime is given by
\begin{equation}
J(\nu)  = C N_0 \nu^{-{{(p-1)}\over 2}} B^{{(p+1)}\over 2}
\label{jnu}
\end{equation}
where $C$ is a constant depending only weakly on $p$. Equation \ref{jnu} implies
that the same observed emissivity can be produced by any combination
of the number density of electrons (scaling as $N_0$) and the
strength of the field $B$. However, total intensity emission does
give important information about the magnitude of any spatial {\it
variation} in field strength (combined with electron density, energy
spectrum etc) given the strong dependence on $B$ of emissivity (eq.\
\ref{jnu}).  The direction of observed polarization in optically
thin regions gives us an (emission-weighted, line-of-sight integrated,
projected) estimate of the local direction of magnetic field, which
is our best probe of the vector properties of the field, while
fractional polarization tells us about its ordering: for a uniform
field the fractional polarization for power-law electron energy
index $p$ is $\Pi = (p+1)/(p + \frac{7}{3})$.

Faraday rotation can complicate
the interpretation of polarization. For the simple case of an external Faraday screen, the
rotation angle $\phi$ is given by
\begin{equation}
\phi = {{c^2}\over{\nu^2}}K\int^s_0 n_{th} {\bf B}.{\rm d}{\bf
s}
\label{faraday}
\end{equation}
where $K$ is a constant, $n_{\rm th}$ is the number density of thermal
electrons, and the integral is along the line of sight to the source.
If the value of $\phi$ varies within the resolution element,
depolarization rather than simple rotation with $\phi \propto
\lambda^2$ will be observed. In some cases, therefore polarized
intensity, and in particular the dependence of fractional polarization
or position angle on frequency, actually tells us more about the
foreground magnetoionic medium (in the host or the Milky Way) than it
does about the source itself. The effects of internal and external
Faraday-active media on polarization are well understood in some
simple analytic cases \citep{Burn66,Jones+ODell77,Cioffi+Jones80}.

Optically thick synchrotron radiation is an essential tool for field
strength measurements on the pc scale; the turnover frequency for
synchrotron self-absorption is sensitively dependent on the magnetic
field strength, with the absorption coefficient going as
$B^{(p+2)/2}$. However, as the turnover also depends on the
normalization of the electron energy spectrum, this method depends on
good information about the structure of the components where the
turnover is observed.

The combination of inverse-Compton and synchrotron emission
gives an excellent constraint on the mean field strength, if both
processes can be measured from the same region; inverse-Compton
emissivity depends only on the number density of electrons (for a known photon
field, often the CMB) so that the field strength may be directly
estimated from the synchrotron emissivity. Again, it is necessary to
know the geometry of the emitting region accurately,

Finally, if all else fails, it is traditional to resort to the
assumptions of equipartition,
\begin{equation}
\frac{B^2}{8\pi} = \int E N(E) {\rm d}E
\end{equation}
or minimum energy \citep{Burbidge56}
\begin{equation}
U_{\rm tot} = \frac{B^2}{8\pi} + \int E N(E) {\rm d}E \quad : \quad
\frac{{\rm d}U_{\rm tot}}{{\rm d} B} = 0
\end{equation}
which, putting in $B$-dependent estimates of $N_0$ from
eq.\ \ref{jnu}, give fairly similar magnetic field values for a given
volume emissivity. Until
recently there has been little observational justification for the equipartition
assumption (though see below) and it should still be applied with caution.

\subsection{Observations: large-scale components}

In the large-scale lobes and plumes of radio galaxies the gold
standard for field strength measurements is the inverse-Compton
technique. Large numbers of powerful (FRII: \citealt{Fanaroff+Riley74})
radio galaxies now have global lobe field strength measurements
\citep{Kataoka+Stawarz05,Croston+05-2} using X-ray inverse-Compton
measurements made with {\it XMM} and {\it Chandra}.  These imply
field strengths close to, but somewhat below, the equipartition
values: a typical field strength in a lobe is of the order 10 $\mu$G.
Fewer such measurements exist for the lower-power FRI radio galaxies,
in part because of their typically brighter thermal environments
but also because some part of their internal pressure is probably
provided by thermal material \citep{Hardcastle+Croston10}. However,
a magnetic field strength around 1 $\mu$G, again close to the
equipartition value, has been estimated for the nearby FRI radio
galaxy Centaurus A based on {\it
  Fermi} $\gamma$-ray detections of its large-scale lobes
\citep{CenA:Abdo+10}.

Inverse-Compton observations also provide important evidence for
the spatial intermittency of the magnetic field: the inverse-Compton
surface brightness (which, for scattering of the CMB, depends only
on the electron energy density and the line-of-sight depth) is seen
to be much more uniform than the synchrotron surface brightness
\citep{Hardcastle+Croston05}. This is as expected if the lobes are
turbulent with an energetically sub-dominant magnetic field, and
similar effects are seen in numerical MHD modeling of lobes
\citep{Tregillis+04,Hardcastle+Krause14}.

Turning to field direction, the apparent field directions inferred
from polarization measurements in the lobes are almost always
perpendicular to the jet far away from the jet termination, with mean
fractional polarization in the range 20--40\% at high frequencies.
This is consistent with the idea that a toroidal component of the
field dominates on large scales, as would be expected if an initially
disordered field expands into the lobes, but compression of a
disordered field can give similar results \citep{Laing80}.

\subsection{Observations: kpc-scale jets and termination features}

Inverse-Compton measurements of field strengths in jets are difficult
because of the existence of \textit{in situ} particle acceleration, which can
give rise to high-energy (e.g. X-ray) synchrotron emission. Arguments
for high bulk speeds in kpc-scale jets, discussed above, assume that
the X-ray emission from these jets is synchrotron, but there is a
degeneracy between the magnetic field strength and the (a priori
unknown) jet speed and angle to the line of sight, so that these
observations are not very useful for measurements. In low-power jets,
where the dynamics are better understood, TeV $\gamma$-ray
observations place some constraints on magnetic field strength
\citep{Hardcastle+Croston11}, implying $B > 20$ $\mu$G to avoid
over-producing the observed TeV emission. These models depend on a
good knowledge of the photon field, in this case the starlight at the
centre of the elliptical host galaxy.

The magnetic field direction is much better understood. In powerful
jets the field direction is almost universally along the jet (although
it is not clear whether the synchrotron emission from the jet is a
reliable tracer of the jet itself --- it may simply show a boundary
layer). In low-power jets, observations of polarization are essential
to models that break the degeneracy between jet speed and angle to the
line of sight \citep{Laing+Bridle02} and when this is done it is found
that the field tends to evolve from a predominantly longitudinal
configuration on the smallest scale to a predominantly toroidal one
after jet deceleration: no significant radial component is required by
the models \citep{Laing+Bridle14}. Helical field models on these
scales for these well-studied low-power jets are ruled out by the lack
of transverse polarization asymmetry.

Powerful jets terminate in compact (1-10 kpc scale), bright features
known as hotspots, which are thought to trace the jet termination
shock. As the downstream material should be subsonic, and so at most
mildly relativistic, hotspots are easier to understand than the jets
themselves.  Their brightness means that they are good sources
of inverse-Compton emission via the synchrotron self-Compton process,
producing X-ray emission \citep{Harris+94}.  Inverse-Compton modeling
implies field strengths of the order 100 $\mu$G in hotspots
of powerful sources \citep{Hardcastle+04,Kataoka+Stawarz05}. Magnetic
field structures in hotspots, as revealed by high-resolution
polarization imaging, are complex, but there is a general tendency
for the projected field direction to be perpendicular to the jet
\citep{Leahy+97}.

\subsection{Observations: pc-scale jets}

Inverse-Compton methods are very uncertain on small scales---high-energy
emission from the jet is often hard to distinguish from emission
from other components of AGN.  In objects where this is not the
case, such as blazars, the jet geometry, bulk speed and angle to
the line of sight are likely to be poorly known. Most methods for
estimating the magnetic field strength on parsec scales rely on
inferences from very long baseline interferometry (VLBI) in the
radio, and hence on synchrotron emission. Typically, the parsec-scale
radio morphology of a bright AGN manifests a one-sided jet structure
due to Doppler boosting, which enhances the emission of the approaching
jet.  As noted above, Lorentz factors on these scales can be high,
$\Gamma \sim 10$--30: constraints come from observations of apparent
motions \citep{Cohetal-07, Listeretal-13} and from radio variability
\citep{Jor-05,Hovetal-09,Savetal-10}.

The apparent base of the jet, which is often the brightest and
least-resolved part, is commonly called the \textit{core}. As noted
above, synchrotron self-absorption, which depends on both the
particle number density $N_{\rm e}$ and the magnetic field $B$,
will affect the appearance of the jet; the core is thought to
represent the jet region where the optical depth is equal to unity.
As the number density and magnetic field depend on distance from
the jet base, the position of the apparent core is expected to
depend on observing frequency. This is the basis of the \textit{core-shift}
method for estimating magnetic field strengths
\citep{Lobanov1998,OSullivanGabuzda2009B}, evaluating the relation between core flux at 
given frequency and mass of black hole \citep{HeinzSun-03}, jet composition 
\citep{Hir-05}, and flow magnetization \citep{NBKZ-15}.  This method necessarily
assumes that the jet is well described by the simple, homogeneous
conical jet models of \cite{BlandfordKonigl1979}. The latter assumption is supported
by multi-frequency observations by \citep{Sok-11}.

As an alternative to the core-shift method, individual components
of the jet can be fitted with self-absorption models
\citep[e.g.][]{Savolainen+08}; this method is less model-dependent
but has larger random errors as a result of the uncertainties on
component size, geometry and Doppler factor. Magnetic field strengths
derived by either method are of the order 1 G on pc scales. Recently,
\cite{2014arXiv1410.7310Z} have proposed a ``hybrid'' method of doing
core-shift analysis which does not rely on the equipartition
assumption, again obtaining similar magnetic field strength values.

On pc scales, the inferred jet magnetic field direction is normally
either parallel or perpendicular to the jet direction
\citep[e.g.][]{ListerHoman2005}, with jets in more powerful objects
tending to have parallel fields. Misaligned polarization that is
definitively not the result of Faraday rotation is rare. The
interpretation of these field structures varies: they are often
attributed to internal jet dynamics, e.g. shocks in the case of
transverse apparent field direction, but could equally well be the
result of an underlying helical magnetic field geometry with varying
pitch angle \citep{LyutikovParievGabuzda2005}. (As noted above,
helical fields are expected in most jet generation models due to the
toroidal component introduced by field winding: see Sections
\ref{sec:intro},\ref{sec:advection}.) As pointed out by
\cite{Blandford1993}, the helical-field model would be supported by
observations of transverse gradients of Faraday rotation across the
jet, provided that there are thermal electrons in the jet or in its
immediate environment to provide the required magnetoionic medium
(eq.\ \ref{faraday}). Detection of such gradients is observationally
difficult, since the jets are poorly resolved, but they have been
observed in a number of objects
\citep[e.g.,][]{Asada2002,OSullivanGabuzda2009RM,CrokeOSullivanGabuzda2010}.
While their interpretation is still controversial
\citep{TaylorZavala2010} these gradients provide at least some direct
evidence for helical field structures in the inner parts of the jets.

\subsection{Core-shifts and jet magnetization}

Two important MHD parameters for describing relativistic flows are the
Michel magnetization parameter $\sigma_{\rm M}$ and the
multiplicity parameter $\lambda$. The first one tells us how strongly the flow
is magnetized at its origin, and it determines the
the maximum possible bulk Lorentz factor of the flow. 
The second one is the dimensionless multiplicity
parameter $\lambda = N_{\rm e}/N_{\rm GJ}$, which is defined as
the ratio of the number density $N_{\rm e}$ to the Goldreich-Julian
(GJ) number density $N_{\rm GJ} = \Omega B/2 \pi c e$ --- the minimum 
number density needed for screening the longitudinal electric field. 
These two parameters are connected with total jet power $P_{\rm jet}$ by
equation~\citep{Beskin-10}
\begin{equation}
\sigma_{\rm M} \approx \frac{1}{\lambda}\left(\frac{P_{\rm
jet}}{P_{\rm A}}\right)^{1/2}. \label{newsigma}
\end{equation}
Here $P_{\rm A} = m_{\rm e}^{2}c^{5}/e^{2} \approx 10^{17}$~erg/s
is the minimum energy loss rate of a central engine which can
accelerate particles to relativistic energies.

There are two theoretical models of plasma
production in a jet.  In the first one, pairs are produced by two-photon collisions
with photons with sufficient energy produced by the inner parts of the accretion disk
~\citep{BZ-77}. In this case we expect \mbox{$\lambda \sim
  10^{10}$--$10^{13}$,} and Michel magnetization parameter
$\sigma_{\rm M} \sim 10$--$10^{3}$. 
The second model of the pair production is a cascade process in non-zero electric field 
in a region with the zero GJ plasma density due to general relativity effects \citep{BIP-92,
HirOk-98}.  
This model gives $\lambda \sim 10^{2}$--$10^{3}$, and magnetization
$\sigma_{\rm M} \sim 10^{10}$--$10^{13}$.
So in both scenarios the flow is strongly magnetized at its base, but 
this large difference in the estimates for the magnetization parameter
$\sigma_{\rm M}$ leads to two different pictures of the
flow structure in jets. Indeed, as has been shown by several authors
\citep{BN-06, Tchekhovskoy-09, Kom-09}, for well-collimated
magnetically dominated MHD jets the Lorentz factors of the particle
bulk motion follows the relation
\begin{equation}
\Gamma \approx r_{\perp}/R_{\rm L}, \label{gamma}
\end{equation}
where $r_{\perp}$ is the distance from the jet axis, and $R_{\rm L} =
c/\Omega$ is the light cylinder radius. The relation~(\ref{gamma})
holds until the flow reaches the equipartition regime --- the Poynting
flux is approximately equal to the particle kinetic energy flux.
Further acceleration is ineffective. For ordinary jets
$r_{\perp}/R_{\rm L} \sim 10^4$--$10^5$. As a result, using the
universal asymptotic solution~(\ref{gamma}), one can find that values
$\sigma_{\rm M} \sim 100$ correspond to the saturation regime when
there is approximately equipartition between the Poynting flux $P_{\rm
  em}$ and the particle kinetic energy flux $P_{\rm part}$. On the
other hand, for $\sigma_{\rm M} \sim 10^{12}$ the jet remains
magnetically dominated ($P_{\rm part} \ll P_{\rm em}$). Thus, the
determination of the Michel magnetization parameter $\sigma_{\rm M}$
is a key point in the analysis of the internal structure of
relativistic jets.

The core-shift measurements described above \citep{Gou-79,
  Mar-83,Sok-11,Pushetal-12} might provide an observational way to
probe $\lambda$ and $\sigma_{\rm M}$. The ideal MHD flow of plasma
with exactly the drift velocity in crossed electric and magnetic
fields does not emit synchrotron radiation. However, some internal
dissipative processes, for example internal shocks and magnetic
reconnection, can produce particles with a non-thermal spectrum
that can account for the observed emission. By measuring the
core-shift discussed in the previous subsection, one can obtain the
combination of magnetic field magnitude $B_{\rm cs}$ and electron
number density $N_{\rm cs}$ in the region with optical depth equal
to unity. On the other hand, $B_{\rm cs}$ and $N_{\rm cs}$ can be
related through the flow magnetization $\sigma=\sigma_{\rm M}/\Gamma-1$
\citep{Tchekhovskoy-09} assuming that the flow is in the saturation
regime, i.e. $\sigma_{\rm cs}\approx\sigma_{\rm M}/2\approx\Gamma_{\rm
cs}$ \citep{BN-06, Tchekhovskoy-09, Kom-09}. Combining the latter
with the definitions of $\lambda$ and $\sigma_{\rm M}$, one can
obtain both $\lambda$ and $\sigma_{\rm M}$ as a function of core-shift
and total jet power $P_{\rm jet}$. The latter can be estimated by
correlating the radio jet luminosity to total power needed to form
the cavities in surrounding gas \citep{Cavagnolo-10}. This analysis
leads to estimates of $\lambda \sim 10^{13}$ and $\sigma\sim 10$
~\citep{NBKZ-15}. The measurements of core-shift by
\citep{Pushetal-12} thus allow the order of Michel magnetization
at the jet base to be estimated, and it can be concluded that the
MHD flow on the observable scales are in the saturation regime,
i.e. there is no effective plasma acceleration by the MHD mechanism
downstream. Moreover, the order of $\sigma_{\rm M}$ correlates well
with the estimated Lorentz factors of jet flows discussed above.

Recently, it has become possible to use the core-shift to measure
the magnetic flux threading the jets and central black holes in a
rather large sample of radio-loud active galactic nuclei
\citep{2014Natur.510..126Z,2014arXiv1410.7310Z}. Dynamically-important
black hole magnetic fields were inferred, indicating that the central
black holes of most if not all radio-loud AGN are in a MAD state
(Section \ref{sec:advection}). MADs have also been inferred in tidal
disruption events \citep{2014MNRAS.437.2744T} and core-collapse
gamma-ray bursts \citep{2014arXiv1409.4414T}. This suggests that
MADs are perhaps not rare or unusual as their name might imply, but
possibly quite the opposite.

\subsection{Observational summary}
\label{sec:observ-summ}
\begin{figure}
\includegraphics[width=\linewidth]{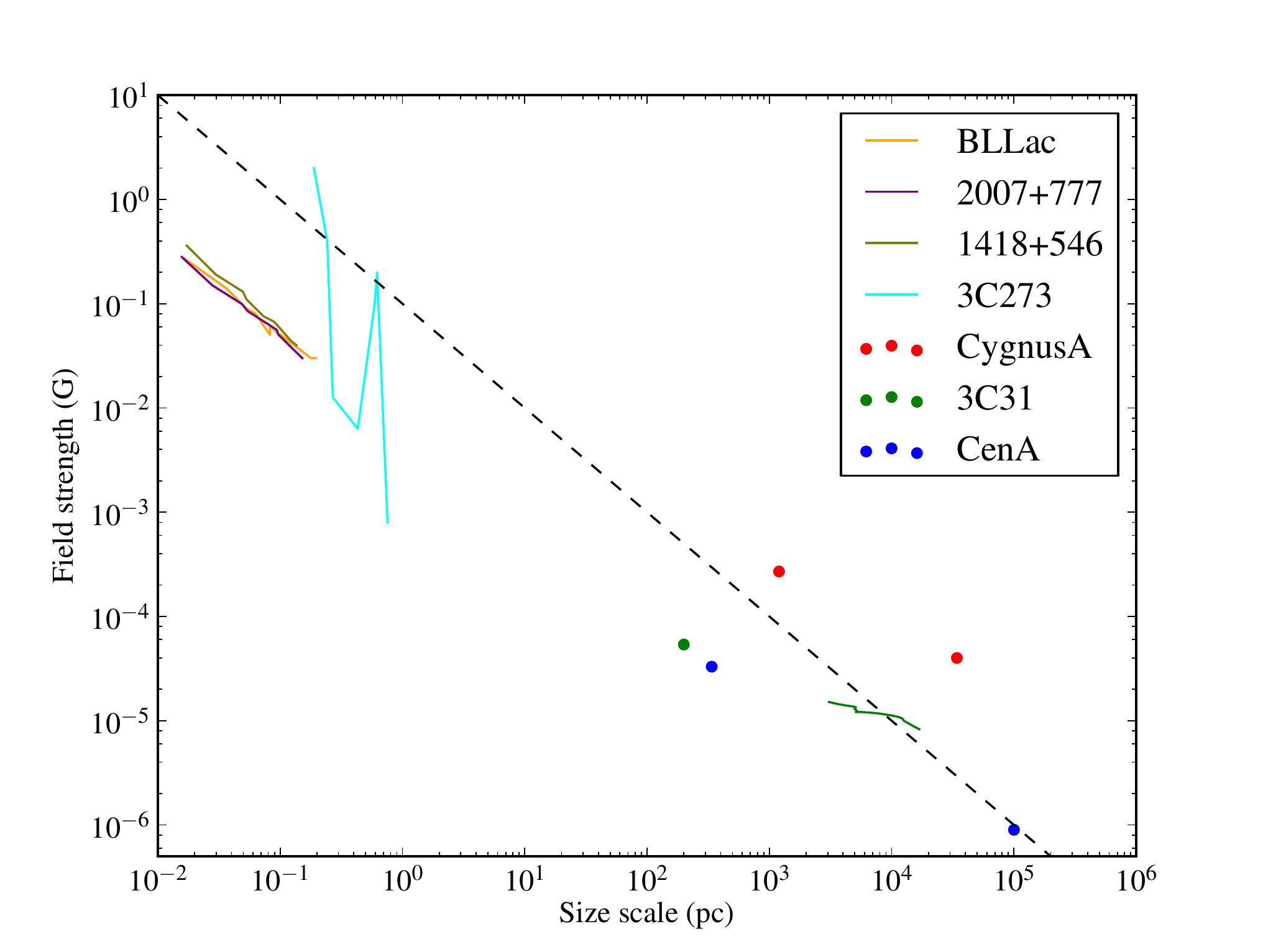}
\caption{Magnetic field strengths on different size scales illustrated using a
  few well-studied objects on pc and kpc scales. The size scale
  plotted here is the transverse radius of the component in which the
  field is measured, not the distance from the nucleus (though in
  general the two scale). Data are taken from
  \cite{OSullivanGabuzda2009B}, \cite{Savolainen+08},
  \cite{Hardcastle+04}, \cite{Hardcastle+Croston10}, 
  \cite{Croston+Hardcastle14},
  \cite{Hardcastle+Croston11} and \cite{CenA:Abdo+10}. The solid line
  shows an arbitrarily normalized line of $B \propto r^{-1}$ --- we
  would expect the normalization of this line to depend on the jet
  power so it is not surprising that all points do not lie on the line.}
\label{magfields}
\end{figure}

To summarize, we now have relatively good information about the
magnetic field strengths on kpc scales, and excellent information
about the vector properties of the field, which, however, are probably
not strongly connected to the field properties at jet generation. On
pc scales and below, magnetic field strengths are not as well
constrained, but such constraints from, e.g., core-shift observations
as we have are consistent with the expectations from models, while
polarization observations are certainly consistent with the presence
of a helical field structure on small scales.

The field strength constraints obtained for some of the objects
discussed above are summarized in Fig.\ \ref{magfields}, which shows a
broad consistency with an $r^{-1}$ scaling over the observable $\sim
7$ orders of magnitude in physical scale. Fields of the order $10^4$
G, and thus magnetic fluxes $\sim 10^{33}$ G
cm$^2$ would be implied by an
extrapolation to scales comparable to the gravitational radius of the
central black hole.  In principle, magnetic fluxes at this level could
be dynamically important (plasma $\beta \sim 1$) and so could affect
the flow of material in the disk as discussed in Section \ref{sec:advection}.

\section{Conclusions}
\label{sec:conclusion}

Accretion disks and the jets they produce are fundamentally magnetic
systems.  A great deal of progress has been made in the theory of
these systems, and in obtaining their properties from observation.  We
began by considering how the entire complex system is broken into 
different components for theoretical analysis.  Where then do we stand
with those components?

First, accretion disks are turbulent, and that turbulence is almost
certainly created by the instability to the MRI. This MRI-induced
turbulence produces internal stress that can transport angular
momentum at the rates required.

Although it seems clear that disks are magnetic, jets appear to
require a more organized magnetic field than is produced by MRI
turbulence. The origin of the jet magnetic field remains uncertain.
A disk dynamo remains a possibility, and dynamo action has been
seen in some simulations, but the field produced is not the type
required for jet launching. In the case of stellar jets, the central
star may also be the source of the required field through its own
internal dynamo.  The possiblity remains that the required field
could be carried in from large radii; however, MRI turbulence does
not seem to accomplish that on its own. More direct infall, taking
place outside the turbulent disk proper, or in a hot, thick accretion
flow with relatively large infall velocities, could work if the
resulting field is able to reconnect as needed to establish the
required topology on the central star or black hole, or through the
central accretion disk.

The general model of jets as magnetized, accelerated and collimated
disk winds is well established today by observations and theoretical
investigations. While for young stars the disk wind seems to play the
major role in jet launching (Blandford-Payne), the jets from AGN and
other relativistic sources are plausibly dominated by processes
energetically supported directly by the spinning black hole
(Blandford-Znajek, \citealt{BZ-77}), and observations, at least on the
smallest scales, show good consistency with the expectations of this
type of model, as discussed in Section \ref{sec:obs}. There are still
many unknowns. For example, the matter content of relativistic jets is
as yet unclear, and in many cases their propagation speed is poorly
constrained; there is no direct measurement of the magnetic field of
jets from stars or of the accretion disk.

Global MHD simulations are able to treat the global jet launching
that is the accretion-ejection process. Usually an initial large-scale
magnetic field has been assumed so far in these models, only a few
simulations were able to consider mean-field disk dynamo.  How such
global, axisymmetric mean-field models match the disk microphysics
and the respective local disk simulations, is still an open question.
The magnetized disks considered in jet launching simulations use a
rather simple model for the (turbulent) magnetic diffusivity or
viscosity and also neglect radiation and radiative transport processes.
However, all these processes seem to be essential for the disk
physics.

Future progress will clearly come from more sophisticated simulation
models, considering the physical effects mentioned above.  The codes
and computational resources may not yet be ready for such an ambitious
goal, as the treatment of the additional physics will require new
numerical methods, and also a higher numerical resolution, together
with a fully 3D treatment.

Simulations have clearly demonstrated the viability of the
Blandford-Znajek type jet powered by a rotating black hole.  Those
simulations further suggest that a dipole field is required, in
addition to the rotating hole, and that the field strength is set
by equipartition with the surrounding pressure (gas, ram, magnetic)
associated with the accretion disk.  The issue of the presence of
such jets comes back to the origin of the axial black hole field,
a problem that remains uncertain, as discussed above.

Finally, we have seen that observations, particularly of AGN jets, are
starting to provide direct tests of jet-launching models through
constraints on the bulk speed, particle content, kinetic power, and
magnetic field strength and configuration of jets. Much progress on at
least some of these issues can be expected from the improvement in
observational capabilities at arcsecond resolutions (and thus
generally kpc scales) to be provided by next-generation radio
facilities such as the upgraded JVLA and LOFAR, and in the future the
SKA. However, the most direct constraints remain those provided by
long-baseline interferometers such as the VLBA and EVN, and continued
support of these is vital if we are to be able to make direct tests of
jet models in the future. 

\begin{acknowledgements}

The authors would like to thank the ISSI team for their support and
hospitality, and for the organization of a great workshop and
fruitful discussions during the week in Bern.  This work was also
supported by NSF grant AST0908869 (JFH) and by funding from the UK
STFC (MJH). AT was supported by NASA through Einstein Postdoctoral
Fellowship grant number PF3-140115 awarded by the {\it Chandra}
X-ray Center, which is operated by the Smithsonian Astrophysical
Observatory for NASA under contract NAS8-03060, and NASA via High-End
Computing (HEC) Program through the NASA Advanced Supercomputing
(NAS) Division at Ames Research Center that provided access to the
Pleiades supercomputer, as well as NSF through an XSEDE computational
time allocation TG-AST100040 on NICS Kraken, Nautilus, TACC Stampede,
Maverick, and Ranch.  AT used Enthought Canopy Python distribution
to generate the figures for this work. JFH wishes to thank Julian Krolik
for useful discussions.

\end{acknowledgements}

% BibTeX users please use one of


\begin{thebibliography}{}



\bibitem[\protect\citeauthoryear{Edlund and Ji}{2014}]{Edlund14}
E. M. Edlund and H. Ji, Phys. Rev. E, \textbf{89}, 021004 (2014)

\bibitem[\protect\citeauthoryear{Abdo et al.}{2010}]{CenA:Abdo+10}
A.A. Abdo et al., Science,  \textbf{328}, 725 (2010)

\bibitem[\protect\citeauthoryear{Anderson et
al.}{2005}]{2005ApJ...630..945A}
 J.M. Anderson, Z.-Y. Li, R. Krasnopolsky, R.D. Blandford, \apj,
 \textbf{630}, 945 (2005).

\bibitem[\protect\citeauthoryear{Asada et al.}{2002}]{Asada2002} 
K. Asada et al., \pasj,  \textbf{54}, L39 (2002)

\bibitem[\protect\citeauthoryear{Avila}{2012}]{Avila12}
M. Avila, Phys. Rev. Lett., \textbf{108}, 124501 (2012)

\bibitem[\protect\citeauthoryear{Bai and Stone}{2013b}]{2013ApJ...769...76B}
X.-N. Bai, J.M. Stone, \apj \textbf{767}, 30 (2013)

\bibitem[\protect\citeauthoryear{Bai and Stone}{2013a}]{2013ApJ...767...30B}
X.-N. Bai,  J.M. Stone, \apj \textbf{769}, 76 (2013)

\bibitem[\protect\citeauthoryear{Balbus and Hawley}{1991}]{BH91} S.A.
Balbus, J.F. Hawley,  \apj, \textbf{376}, 214 (1991)

\bibitem[\protect\citeauthoryear{Balbus and Hawley}{1998}]{BH98}
S.A. Balbus,  J.F. Hawley, Rev. Mod. Phys., \textbf{70},
1 (1998) 

\bibitem[\protect\citeauthoryear{Balbus, Hawley and
Stone}{1996}]{BHS96} S.A. Balbus, J.F. Hawley, J.M. Stone,
\apj, \textbf{467}, 76 (1996)

\bibitem[\protect\citeauthoryear{Balbus and Papaloizou}{1999}]{BalPap99}
S.A. Balbus, J.C.B. Papaloizou, \apj, \textbf{521},
650 (1990)

\bibitem[\protect\citeauthoryear{Beckwith, Hawley and Krolik}{2008}]{bhk08} 
K.R.C. Beckwith, J.F. Hawley,  J.H. Krolik, \apj,
\textbf{678}, 1180  (2008)

\bibitem[\protect\citeauthoryear[{Beckwith, Hawley and Krolik}{2009}]{bhk09}
K.R.C. {Beckwith}, J.F. {Hawley}, J.H. {Krolik}, \apj,
\textbf{707}, 428 (2009)

\bibitem[\protect\citeauthoryear{Beskin}{2010}]{Beskin-10}
V.S.~Beskin, \mbox{Physics-Uspekhi} \textbf{53}, 1199 (2010)

\bibitem[\protect\citeauthoryear{Beskin and Nokhrina}{2006}]{BN-06}
V.S. Beskin, E.E.~Nokhrina, \mnras, \textbf{367}, 375
(2006)

\bibitem[\protect\citeauthoryear{Beskin et al.}{1992}]{BIP-92}
V.S. Beskin, Ya.N. Istomin, V.I. Pariev, Sov.~Astron., \textbf{36}, 642
(1992)

\bibitem[\protect\citeauthoryear{Best and
Heckman}{2012}]{Best+Heckman12} P.N. Best,  T.M. Heckman, 
\mnras,  \textbf{421}, 1569 (2012)

\bibitem[\protect\citeauthoryear{Bicknell}{1994}]{Bicknell94} 
G.V. Bicknell, \apj, \textbf{422}, 542 (1994)

\bibitem[\protect\citeauthoryear{Blandford}{1993}]{Blandford1993} 
R.D. Blandford, in \textit{Astrophysics and Space Science
Library}, \textbf{103}, 15 (1993)

\bibitem[\protect\citeauthoryear{Blandford and
K\"onigl}{1979}]{BlandfordKonigl1979}
R.D. Blandford, A. K\"onigl, \apj, \textbf{232}, 34 (1979)

\bibitem[\protect\citeauthoryear{Blandford and Payne}{1982}]{BP82}
R.D. Blandford,  D.G. Payne, \mnras, \textbf{199}, 883 (1982)

\bibitem[\protect\citeauthoryear{Blandford and Znajek}{1977}]{BZ-77}
R.D. Blandford, R.L. Znajek, \mnras, \textbf{179}, 433 (1977)

\bibitem[\protect\citeauthoryear{Brandenburg et al.}{1995}]{Brandenburg95}
A. Brandenburg, A. Nordlund, R. F. Stein, U. Torkelsson, \apj, \textbf{446}, 741 (1995)

\bibitem[\protect\citeauthoryear{Burbidge}{1956}]{Burbidge56}
G. Burbidge, \apj, \textbf{124}, 416 (1956)

\bibitem[\protect\citeauthoryear{Burn}{1966}]{Burn66} 
B.J.~Burn, \mnras, \textbf{133}, 67 (1966)

\bibitem[\protect\citeauthoryear{Carrasco-Gonz\'alez et
al.}{2010}]{2010Sci...330.1209C} C. Carrasco-Gonz\'alez, L.F.
Rodr\'iguez, G. Anglada, J. Mart\'i, J.M. Torrelles,  M. Osorio,
Science, \textbf{179}, 433 (2010).

\bibitem[\protect\citeauthoryear{Casse and
Ferreira}{2000}]{2000A&A...361.1178C} F. Casse, J. Ferreira, \aap,
\textbf{361}, 1178 (2000).

\bibitem[\protect\citeauthoryear{Casse and
Keppens}{2002}]{2002ApJ...581..988C} F. Casse, R. Keppens, \apj,
\textbf{581}, 988 (2002).

\bibitem[\protect\citeauthoryear{Casse and
Keppens}{2004}]{2004ApJ...601...90C} F. Casse, R. Keppens, \apj,
\textbf{601}, 90 (2004).

\bibitem[\protect\citeauthoryear{Cavagnolo et~al.}{2010}]{Cavagnolo-10}
K.W. Cavagnolo, B.R. McNamara, P.E.J. Nulsen et al., \apj, \textbf{720}, 1066 (2010)

\bibitem[\protect\citeauthoryear{Cioffi and
Jones}{1980}]{Cioffi+Jones80} D.F. Cioffi, T.W. Jones,  \aj,
\textbf{85}, 368 (1980)

\bibitem[\protect\citeauthoryear{Cohen et al.}{2007}]{Cohetal-07}
M.H. Cohen et al., \apj, \textbf{658}, 232 (2007)

\bibitem[\protect\citeauthoryear{Contopoulos}{1994a}]{1994ApJ...429..139C}
J. Contopoulos, R.V.E. Lovelace, \apj, \textbf{429}, 139 (1994a).

\bibitem[\protect\citeauthoryear{Contopoulos}{1994b}]{1994ApJ...432..508C}
J. Contopoulos,  \apj, \textbf{432}, 508 (1994b)

\bibitem[\protect\citeauthoryear{Croke, O'Sullivan and
Gabuzda}{2010}]{CrokeOSullivanGabuzda2010} S.M. Croke,  S.P. O'Sullivan,  
 D.C. Gabuzda, \mnras, \textbf{402}, 259 (2010)

\bibitem[\protect\citeauthoryear{Croston et al.}{2005}]{Croston+05-2} 
J.H. Croston, et al., \apj, \textbf{626}, 733  (2005)

\bibitem[\protect\citeauthoryear{Croston and
Hardcastle}{2014}]{Croston+Hardcastle14} J.H. Croston, 
M.J. Hardcastle,  \mnras,  \textbf{438}, 3310 (2014)

\bibitem[\protect\citeauthoryear{De Villiers, Hawley, Krolik and
Hirose}{2005}]{dVhkh05} J.-P. De~Villiers, J.F. Hawley, J.H. Krolik, 
S. Hirose,  \apj,  \textbf{620}, 878 (2005)

\bibitem[\protect\citeauthoryear{Donati et al.}{2005}]{2005Natur.438..466D} 
J.-F. Donati, F. Paletou, J. Bouvier, J. Ferreira, \nat, \textbf{438}, 466 (2005)

\bibitem[\protect\citeauthoryear{Fanaroff and
Riley}{1974}]{Fanaroff+Riley74} B.L. Fanaroff,  J.M. Riley,
\mnras, \textbf{167}, 31P (1974)

\bibitem[\protect\citeauthoryear{Fendt}{2006}]{2006ApJ...651..272F}
Ch. Fendt, \apj, \textbf{651}, 272 (2006).

\bibitem[\protect\citeauthoryear{Fendt}{2009}]{2009ApJ...692..346F}
Ch. Fendt, \apj, \textbf{692}, 346 (2009).

\bibitem[\protect\citeauthoryear{Fendt et al.}{1995}]{1995A&A...300..791F}
Ch. Fendt, M. Camenzind, S. Appl, \aap, \textbf{300}, 791 (1995).

\bibitem[\protect\citeauthoryear{Fendt and Cemeljic}{2002}]{2002A&A...395.1045F}
Ch. Fendt, and M. Cemeljic, \aap, \textbf{395}, 1045 (2002).

\bibitem[\protect\citeauthoryear{Fendt and Zinnecker}{1998}]{1998A&A...334..750F}
Ch. Fendt, and H. Zinnecker, \aap, \textbf{334}, 750 (1998).

\bibitem[\protect\citeauthoryear{Fendt and
Sheikhnezami}{2013}]{2013ApJ...774...12F} Ch. Fendt, S. Sheikhnezami,
\apj, \textbf{774}, 12 (2013).

\bibitem[Fernandes et al.(2011)]{2011MNRAS.411.1909F} C.A.C. Fernandes, 
M.J. Jarvis, S.  Rawlings, et al., \mnras, \textbf{411}, 1909 (2011)

\bibitem[\protect\citeauthoryear{Ferreira}{1997}]{1997A&A...319..340F}
J. Ferreira, \aap, \textbf{319}, 340 (1997).

\bibitem[\protect\citeauthoryear{Ferreira and
Pelletier}{1993}]{1993A&A...276..625F} J. Ferreira, G. Pelletier,
\aap, \textbf{276}, 625 (1993).

\bibitem[\protect\citeauthoryear{Frank et al.}{2014}]{2014prpl.conf..451F}
A. Frank et al.  in \textit{Protostars and Planets VI}
H. Beuther, R.S. Klessen, C.P. Dullemond, T. Henning (eds.), 
(University of Arizona Press, Tucson 2014)

\bibitem[\protect\citeauthoryear{Fromang 
and Stone}{2009}]{2009A&A...507...19F} S. Fromang, S., J.M. Stone,
\aap, \textbf{507}, 19 (2009)

\bibitem[\protect\citeauthoryear{Gammie, Shapiro and McKinney}{2004}]{gsm04} 
C.F. Gammie,  S.L. Shapiro, J.C. McKinney, \apj,
\textbf{602}, 312 (2004)

\bibitem[Ghisellini et al.(2010)]{2010MNRAS.402..497G} G. Ghisellini,
F. Tavecchio, L. Foschini, L., et al., \mnras, \textbf{402}, 497 (2010)

\bibitem[Ghisellini et al.(2014)]{2014Nat..515...376} G. Ghisellini,
F. Tavecchio, L. Maraschi, A. Celotti and T. Sbarrato, \nat,
\textbf{515}, 376 (2014)

\bibitem[\protect\citeauthoryear{Gould}{1979}]{Gou-79}
R.J. Gould, \aap, \textbf{76}, 306 (1979)

\bibitem[\protect\citeauthoryear{Gressel}{2010}]{Gressel10}
O. Gressel, \mnras, \textbf{405}, 41 (2010)

\bibitem[\protect\citeauthoryear{Guan and Gammie}{2009}]{GG09}
X. Guan, C.F. Gammie, \apj, \textbf{697}, 1901 (2009)

\bibitem[\protect\citeauthoryear{Guan and Gammie}{2011}]{GG11}
X. Guan, C.F. Gammie, \apj, \textbf{728}, 130 (2011)

\bibitem[\protect\citeauthoryear{Hardcastle}{2006}]{Hardcastle06} 
M.J. Hardcastle, \mnras,  \textbf{366}, 1465 (2006)

\bibitem[\protect\citeauthoryear{Hardcastle and
Croston}{2005}]{Hardcastle+Croston05} M.J. Hardcastle, J.H. Croston,
 \mnras, \textbf{363}, 649 (2005)

\bibitem[\protect\citeauthoryear{Hardcastle and
Croston}{2010}]{Hardcastle+Croston10} M.J. Hardcastle,  
J.H. Croston,  \mnras, \textbf{404}, 2018 (2010)

\bibitem[\protect\citeauthoryear{Hardcastle and
Croston}{2011}]{Hardcastle+Croston11} M.J. Hardcastle, 
J.H. Croston, \mnras, \textbf{415}, 133 (2011)

\bibitem[\protect\citeauthoryear{Hardcastle, Croston and
Kraft}{2007}]{Hardcastle+07} M.J. Hardcastle, J.H. Croston,
R.P.  Kraft, \apj, \textbf{669}, 893 (2007)

\bibitem[\protect\citeauthoryear{Hardcastle, Evans and
Croston}{2009}]{Hardcastle+09} M.J. Hardcastle,  D.A. Evans, 
J.H. Croston, \mnras,   \textbf{396}, 1929 (2009)

\bibitem[\protect\citeauthoryear{Hardcastle, Harris, Worrall and
Birkinshaw}{2004}]{Hardcastle+04} M.J. Hardcastle,  D.E. Harris,
D.M. Worrall, M. Birkinshaw,  \apj, \textbf{612}, 729 (2004)

\bibitem[\protect\citeauthoryear{Hardcastle and
Krause}{2014}]{Hardcastle+Krause14} M.J. Hardcastle,  M.G. Krause,
\mnras,  \textbf{443}, 1482 (2014)

\bibitem[\protect\citeauthoryear{Harris, Carilli and
Perley}{1994}]{Harris+94} D.E. Harris, C.L. Carilli, R.A. Perley,
\nat, \textbf{367}, 713 (1994)

\bibitem[\protect\citeauthoryear{Hawley}{2000}]{Hawley:2000}
J.F. Hawley \apj, \textbf{528}, 462 (2000)

\bibitem[\protect\citeauthoryear{Hawley}{2002}]{Hawley:2002}
J.F. Hawley, S.A. Balbus, \apj, \textbf{573}, 738 (2002)

\bibitem[\protect\citeauthoryear{Hawley, Gammie and Balbus}{1995}]{HGB95} 
J.F. Hawley,  C.F. Gammie,  S.A. Balbus, \apj,
\textbf{440}, 742 (1995)

\bibitem[\protect\citeauthoryear{Hawley, Gammie and Balbus}{1996}]{HGB96} 
J.F. Hawley,  C.F. Gammie,  S.A. Balbus, \apj,
\textbf{464}, 690 (1996)

\bibitem[\protect\citeauthoryear{Hawley and Krolik}{2006}]{hk06}
J.F. Hawley, J.H. Krolik, \apj, \textbf{641}, 103 (2006)

\bibitem[\protect\citeauthoryear{Heinz and Sunyaev}{2003}]{HeinzSun-03}
S.~Heinz, R.~Sunyaev, \mnras, \textbf{343}, L59-L64 (2003)

\bibitem[\protect\citeauthoryear{Hester et al.}{2002}]{2002ApJ...577L..49H}
J.J. Hester, K. Mori, D. Burrows, J.S. Gallagher, J.R. Graham, M. Halverson, A. Kader, F.C. Michel, P. Scowen,
\apj, \textbf{577}, L49 (2002)

\bibitem[\protect\citeauthoryear{Hirotani and Okamoto}{1998}]{HirOk-98}
K. Hirotani, I. Okamoto, \apj, \textbf{497}, 563 (1998)

\bibitem[\protect\citeauthoryear{Hirotani}{2005}]{Hir-05}
K. Hirotani, \apj, \textbf{619}, 73 (2005)

\bibitem[\protect\citeauthoryear{Hovatta et~al.}{2009}]{Hovetal-09}
T. Hovatta, E. Valtaoja, M. Tornikoski, A. L\"ahteenm\"aki, 
\aap, \textbf{498}, 723 (2009)

\bibitem[\protect\citeauthoryear{Ji et al.}{2006}]{Ji06}
H. Ji, M. Burin, E. Schartman, J. Goodman, \nat,  \textbf{444}, 343 (2006)

\bibitem[\protect\citeauthoryear{Jones and O'Dell}{1977}]{Jones+ODell77}
T.W. Jones, S.L. O'Dell,  \apj,  \textbf{214}, 522 (1977)

\bibitem[\protect\citeauthoryear{Jorstad et~al.}{2005}]{Jor-05}
S.G. Jorstad, et al., \aj, \textbf{130}, 1418 (2005)

\bibitem[\protect\citeauthoryear{Kataoka and
Stawarz}{2005}]{Kataoka+Stawarz05} J. Kataoka,  L.  Stawarz,
\apj, \textbf{622}, 797 (2005)
 
\bibitem[\protect\citeauthoryear{Kato et al.}{2002}]{2002ApJ...565.1035K}
S.X. Kato, T. Kudoh, K. Shibata, \apj, \textbf{565}, 1035 (2002).

\bibitem[\protect\citeauthoryear{Keppens et al.}{2008}]{2008A&A...486..663K}
R. Keppens, Z. Meliani, B. van der Holst, F. Casse, \aap, \textbf{486}, 663 (2008).

\bibitem[\protect\citeauthoryear{K\"ording et
    al.}{2008}]{2008Sci...320.1318K}K\"ording, E., Rupen, M., Knigge,
  C., Fender, R., Dhawan, V., Templeton, M., Muxlow, T., 2008, Science,
  \textbf{320}, 1318 (2008)

\bibitem[\protect\citeauthoryear{K\"ording et
    al.}{2011}]{2011MNRAS.418L.129K}K\"ording, E., Knigge, C.,
  Tzioumis, T., Fender, R., \mnras, \textbf{418}, L129 (2011).

\bibitem[\protect\citeauthoryear{Komissarov et al.}{2009}]{Kom-09}
S.S. Komissarov, N. Vlahakis, A. K\"onigl, M. V. Barkov~M.~V.,
\mnras, \textit{394}, 1182 (2009)

\bibitem[\protect\citeauthoryear{Konar and
Hardcastle}{2013}]{Konar+Hardcastle13} C. Konar, M.J. Hardcastle,
\mnras,  \textbf{395}, 457 (2013)

\bibitem[\protect\citeauthoryear{K\"onigl et al.}{2010}]{2010MNRAS.401..479K}
A. K\"onigl, R. Salmeron, M. Wardle, \mnras, \textbf{401}, 479 (2010).

\bibitem[\protect\citeauthoryear{Krasnopolsky et
al.}{1999}]{1999ApJ...526..631K} R. Krasnopolsky, Z.-Y. Li, R.D.
Blandford,  \apj, \textbf{526}, 631 (1999).

\bibitem[\protect\citeauthoryear{Kudoh et al.}{1998}]{1998ApJ...508..186K}
T. Kudoh, R. Matsumoto, K. Shibata, \apj, \textbf{508}, 186 (1998).

\bibitem[\protect\citeauthoryear{Kuwabara et al.}{2005}]{2005ApJ...621..921K}
T. Kuwabara, K. Shibata, T. Kudoh, R. Matsumoto, \apj, \textbf{621}, 921 (2005).

\bibitem[\protect\citeauthoryear{Kuwabara et al.}{2005}]{2013A&A...550A..61L}
G. Lesur, J. Ferreira, G. Ogilvie, \aap, \textbf{550}, 61 (2013).

\bibitem[\protect\citeauthoryear{Laing}{1980}]{Laing80} R.A. Laing,
\mnras, \textbf{193}, 439 (1980)

\bibitem[\protect\citeauthoryear{Laing and
Bridle}{2002}]{Laing+Bridle02} R.A. Laing, A.H. Bridle,
\mnras,  \textbf{336}, 328 (2002)

\bibitem[\protect\citeauthoryear{Laing and
Bridle}{2014}]{Laing+Bridle14} R.A. Laing, A.H. Bridle,
\mnras,  \textbf{437}, 3405 (2014)

\bibitem[\protect\citeauthoryear{Leahy et al.}{1997}]{Leahy+97} 
J.P. Leahy, et al., \mnras, \textbf{291}, 20 (1997)

\bibitem[\protect\citeauthoryear{Lesur 
and Longaretti}{2009}]{2009A&A...504..309L} G. Lesur, P.-Y. Longaretti, 
\aap, \textbf{504}, 309 (2009)

\bibitem[\protect\citeauthoryear{Lesur and Longaretti}{2007}]{Lesur07} 
G. Lesur, P.-Y. Longaretti, \mnras, \textbf{378}, 1471 (2007)

\bibitem[\protect\citeauthoryear{Lesur, Ferreira and Ogilvie }{2013}]{Lesur13} 
G. Lesur, J. Ferreira and G. I. Ogilvie, \aa, \textbf{550}, A61 (2013)

\bibitem[\protect\citeauthoryear{Li}{1993}]{1993ApJ...415..118L}
Z.-Y. Li, \apj, \textbf{415}, 118 (1993).

\bibitem[\protect\citeauthoryear{Li}{1995}]{1995ApJ...444..848L}
Z.-Y. Li, \apj, \textbf{444}, 848 (1995).

\bibitem[\protect\citeauthoryear{Lister and
Homan}{2005}]{ListerHoman2005} M.L. Lister,  D.C. Homan,
\aj, \textbf{130}, 1389 (2005)

\bibitem[\protect\citeauthoryear{Lister et al.}{2009a}]{Lister+09}
M.L. Lister, et al., \aj, \textbf{138}, 1874 (2009)

\bibitem[\protect\citeauthoryear{Lister et al.}{2013}]{Listeretal-13} 
M.L. Lister, et al., \aj, \textbf{146}, 120 (2013) 

\bibitem[\protect\citeauthoryear{Lobanov}{1998}]{Lobanov1998}
A.P. Lobanov,  \aap, \textbf{330}, 79 (1998)

\bibitem[\protect\citeauthoryear{Lovelace et
al.}{2010}]{2010MNRAS.408.2083L} R.V.E. Lovelace, M.M. Romanova,
G.V. Ustyugova, A.V. Koldoba, \mnras, \textbf{408}, 2083 (2010)

\bibitem[Lubow et al.(1994)]{1994MNRAS.267..235L} S.H. Lubow, J.C.B. Papaloizou,
J.E. Pringle, \mnras, \textbf{267}, 235  (1994)

\bibitem[\protect\citeauthoryear{Lynden-Bell}{1969}]{LyndenBell69}
D. Lynden-Bell,  \nat, \textbf{223}, 690 (1969)

\bibitem[\protect\citeauthoryear{Lynden-Bell and Pringle}{1974}]{Lynden74} 
D. Lynden-Bell, J.E.  Pringle, \mnras, \textbf{168}, 603 (1974)

\bibitem[\protect\citeauthoryear{Lyutikov, Pariev and
Gabuzda}{2005}]{LyutikovParievGabuzda2005} M. Lyutikov, V.I. Pariev,
D.C. Gabuzda, \mnras,  \textbf{360}, 869 (2005)

\bibitem[\protect\citeauthoryear{Marscher}{1983}]{Mar-83}
A.P. Marscher,  \apj, \textbf{264}, 296 (1983)

\bibitem[\protect\citeauthoryear{Marshall et al.}{2011}]{Marshall+11}
H.L. Marshall, et al., \apjs,  \textbf{193}, 15 (2011)

\bibitem[\protect\citeauthoryear{McKinney}{2005}]{McKinney:2005}
J.C. McKinney,  \apj, \textbf{630}, L5 (2005)

\bibitem[\protect\citeauthoryear{McKinney}{2006}]{McKinney:2006}
J.C. McKinney, \mnras, \textbf{368}, 1561 (2006)

\bibitem[\protect\citeauthoryear{McKinney and Gammie}{2004}]{mg04} 
J.C. McKinney, C.F. Gammie, \apj, \textbf{611}, 977 (2004)

\bibitem[McKinney et al.(2012)]{2012MNRAS.423.3083M} J.C. McKinney, 
A. Tchekhovskoy, R.D. Blandford, \mnras, \textbf{423}, 3083 (2012) 

\bibitem[\protect\citeauthoryear{McNamara et al.}{2009}]{2009ApJ...698..594M} 
B.R. McNamara,
F. Kazemzadeh, D.A. Rafferty, et al., \apj, \textbf{698}, 594 (2009)

\bibitem[McNamara et al.(2011)]{2011ApJ...727...39M} B.R. McNamara,
M. Rohanizadegan, P.E.J. Nulsen, \apj, \textbf{727}, 39 (2011)

\bibitem[\protect\citeauthoryear{Meisenheimer, Yates and
R\"oser}{1997}]{Meisenheimer+97} K. Meisenheimer,  M. G. Yates,
H.-J. R\"oser,  \aap, \textbf{325}, 57 (1997)

\bibitem[\protect\citeauthoryear{Meliani et al.}{2006}]{2006A&A...460....1M}
Z. Meliani, F. Casse, C. Sauty, \aap, \textbf{460}, 1 (2006).

\bibitem[\protect\citeauthoryear{Mignone et al.}{2010}]{2010MNRAS.402....7M}
A. Mignone, P. Rossi, G. Bodo, A. Ferrari, S. Massaglia, \mnras, \textbf{402}, 7 (2010).

\bibitem[\protect\citeauthoryear{Mingo et al.}{2014}]{Mingo+14} 
B. Mingo, et al., \mnras, \textbf{440} 269 (2014)

\bibitem[\protect\citeauthoryear{Mullin and
Hardcastle}{2009}]{Mullin+Hardcastle09} L.M. Mullin,
M.J. Hardcastle,  \mnras, \textbf{398}, 1989 (2009)

\bibitem[\protect\citeauthoryear{Murphy et al.}{2010}]{2010A&A...512A..82M}
G.C. Murphy, J. Ferreira,  C. Zanni, \aap, \textbf{512}, 82 (2010).

\bibitem[Narayan et al.(2003)]{nia03} R. Narayan, 
I.V. Igumenshchev, M. A. Abramowicz, \pasj, \textbf{55}, L69 (2003) 

\bibitem[Narayan et al.(2012)]{2012MNRAS.426.3241N} R. Narayan, 
A. S{\"a}dowski, R.F. Penna, A.K. Kulkarni, \mnras, \textbf{426},
3241 (2012)

\bibitem[Nemmen 
and Tchekhovskoy(2014)]{2014arXiv1406.7420N} R.S. Nemmen, A. Tchekhovskoy, arXiv:1406.7420 (2014)

\bibitem[\protect\citeauthoryear{Nokhrina et al.}{2015}]{NBKZ-15}  
E.E. Nokhrina, V.S. Beskin, Y.Y. Kovalev, A.A. Zheltoukhov,
\mnras, \textbf{447}, 2726 (2015)

\bibitem[\protect\citeauthoryear{Norman et al.}{1982}]{NWSS82}  
M.L. Norman, K.-H.A. Winkler, L.L. Smarr, M.D. Smith,
\aap, \textbf{113}, 285 (1982)

\bibitem[\protect\citeauthoryear{O'Sullivan and
Gabuzda}{2009a}]{OSullivanGabuzda2009RM} 
S. P. O'Sullivan, D. C. Gabuzda,
\mnras, \textbf{393}, 429 (2009a)

\bibitem[\protect\citeauthoryear{O'Sullivan and
Gabuzda}{2009b}]{OSullivanGabuzda2009B}
S. P. O'Sullivan, D. C. Gabuzda,
\mnras, \textbf{400}, 26 (2009b)

\bibitem[\protect\citeauthoryear{Ouyed and Pudritz}{1997}]{1997ApJ...482..712O}
R. Ouyed, R.E. Pudritz, \apj, \textbf{482}, 710 (1997).

\bibitem[\protect\citeauthoryear{Ouyed et al.}{2003}]{2003ApJ...582..292O}
R. Ouyed, D.A. Clarke, R.E. Pudritz, \apj, \textbf{582}, 292 (2003).

\bibitem[\protect\citeauthoryear{Paoletti and Lathrop}{2011}]{Paoletti11}
M. S. Paoletti and D. P. Lathrop, Phys. Rev. Lett., \textbf{106}, 024501 (2011)

\bibitem[\protect\citeauthoryear{Pavlov et al.}{2003}]{2003ApJ...591.1157P} 
G.G. Pavlov, M.A. Teter, O. Kargaltsev, D. Sanwal, \apj, \textbf{591}, 1157 (2003)

\bibitem[\protect\citeauthoryear{Pelletier and
Pudritz}{1992}]{1992ApJ...394..117P} G. Pelletier, R.E. Pudritz,
\apj, \textbf{394}, 117 (1992).

\bibitem[\protect\citeauthoryear{Porth and Fendt}{2010}]{2010ApJ...709.1100P}
O. Porth, Ch. Fendt, \apj, \textbf{709}, 1100 (2010).

\bibitem[\protect\citeauthoryear{Porth et al.}{2011}]{2011ApJ...737...42P}
O. Porth, Ch. Fendt,  Z. Meliani, B. Vaidya, \apj, \textbf{709}, 1100 (2011).

\bibitem[\protect\citeauthoryear{Porth}{2013}]{2013MNRAS.429.2482P}
O. Porth, \mnras, \textbf{429}, 2482 (2013).

\bibitem[\protect\citeauthoryear{Proga and Kallman}{2004}]{2004ApJ...616..688P}
D. Proga, T.R. Kallman, \apj, \textbf{616}, 688 (2004).

\bibitem[\protect\citeauthoryear{Proga et al.}{2000}]{2000ApJ...543..686P}
D. Proga, J.M. Stone, T.R. Kallman, \apj, \textbf{543}, 686 (2000).

\bibitem[\protect\citeauthoryear{Pudritz, Hardcastle and
Gabuzda}{2012}]{Pudritz+12}  R.E. Pudritz,  M.J. Hardcastle, 
D.C. Gabuzda, \ssr, \textbf{169}, 27 (2012)

\bibitem[\protect\citeauthoryear{Pudritz and Norman}{1983}]{1983ApJ...274..677P}
R.E. Pudritz, C.A. Norman, \apj, \textbf{274}, 677 (1983).

\bibitem[\protect\citeauthoryear{Pudritz and Norman}{2006}]{2006MNRAS.365.1131P}
R.E. Pudritz, C. Rogers, R. Ouyed, \mnras, \textbf{365}, 1131 (2006).

\bibitem[\protect\citeauthoryear{Pudritz et al.}{2007}]{2007prpl.conf..277P} 
R.E. Pudritz, et al.  in \textit{Protostars and Planets V} 
B. Reipurth, D. Jewitt, K.  Keil (eds.), 
(University of Arizona Press, Tucson, 2007).

\bibitem[\protect\citeauthoryear{Pushkarev et al.}{2012}]{Pushetal-12}
A. B. Pushkarev, T. Hovatta, Y. Y. Kovalev, et al.,\aap, \textbf{545}, A113 (2012)

\bibitem[\protect\citeauthoryear{Ramsey and Clarke}{2011}]{2011ApJ...728L..11R}
J.P. Ramsey, D.A. Clarke, \apj, \textbf{728}, 11 (2011).

\bibitem[\protect\citeauthoryear{Rawlings and
Saunders}{1991}]{Rawlings+Saunders91} S. Rawlings, R. Saunders,
\nat, \textbf{349}, 138 (1991)

\bibitem[\protect\citeauthoryear{Rees et al.}{1982}]{Rees:1982} M.J.
Rees, M.C. Begeman, R.D. Blandford, E.S. Phinney, \nat,
\textbf{295}, 17 (1982)

\bibitem[\protect\citeauthoryear{Salmeron et al.}{2011}]{2011MNRAS.412.1162S}
R. Salmeron, A. K{\"o}nigl,  M. Wardle, \mnras, \textbf{412}, 1163 (2011).

\bibitem[\protect\citeauthoryear{Savolainen, Wiik, Valtaoja
and Tornikoski}{2008}]{Savolainen+08} T. Savolainen, K. Wiik,
E. Valtaoja, M. Tornikoski, in \textit{Extragalactic jets:
Theory and Observation from Radio to Gamma Ray}, T. A. Rector and D. S.
De Young (eds), 451 (2008)

\bibitem[\protect\citeauthoryear{Savolainen et al.}{2010}]{Savetal-10}
T.~Savolainen, D.C.Homan, T.~Hovatta, M.~Kadler, Y.Y.~Kovalev,
M.L.~Lister, E.~Ros, J.A.~Zensus, \aap, \textbf{512}, A24 (2010)

\bibitem[\protect\citeauthoryear{Sauty and Tsinganos}{1994}]{1994A&A...287..893S}
C. Sauty, K. Tsinganos, \aap, \textbf{287}, 893 (1994).

\bibitem[\protect\citeauthoryear{Schartman et al.}{2012}]{Schartman12}
E. Schartman, H. Ji, M. Burin, and J. Goodman, \aa, \textbf{543}, A94 (2012)

\bibitem[\protect\citeauthoryear{Scheuer}{1974}]{Scheuer74} 
P. A. G. Scheuer, \mnras,  \textbf{166}, 513 (1974)

\bibitem[\protect\citeauthoryear{Shakura and
Sunyaev}{1973}]{Shakura73} N. I. Shakura, R. A. Sunyaev, 
\aap, \textbf{24}, 337 (1973)

\bibitem[\protect\citeauthoryear{Shang et al.}{2007}]{2007prpl.conf..261S}
H. Shang, Z.-Y. Li, N. Hirano,
in \textit{Protostars and Planets V},
B. Reipurth, D. Jewitt, and K. Keil (eds.), 
(University of Arizona Press, Tucson, 2007).

\bibitem[\protect\citeauthoryear{Sheikhnezami et
al.}{2012}]{2012ApJ...757...65S} S. Sheikhnezami, Ch. Fendt, O.
Porth, B. Vaidya, J. Ghanbari, \apj, \textbf{757}, 65 (2012).

\bibitem[\protect\citeauthoryear{Shibata and Uchida}{1985}]{SU85}
K. Shibata, Y. Uchida, \pasj, \textbf{37}, 31  (1985)

\bibitem[\protect\citeauthoryear{Simon and Hawley}{2009}]{Simon09b}
J. B. Simon, J. F. Hawley, \apj, \textbf{707}, 833 (2009)

\bibitem[\protect\citeauthoryear{Simon, Hawley and Beckwith}{2009}]{Simon09}
J. B. Simon, J. F. Hawley, K. Beckwith, \apj, \textbf{690}, 974 (2009)

\bibitem[\protect\citeauthoryear{Soker and Lasota}{2004}]{2004A&A...422.1039S}
N. Soker, J.P. Lasota, \aa, \textbf{422}, 1039 (2004)

\bibitem[\protect\citeauthoryear{Sokolovsky et al.}{2011}]{Sok-11}
K. V. Sokolovsky, et al, \aap, \textbf{532}, A38 (2011)

\bibitem[\protect\citeauthoryear{Stepanovs and Fendt}{2014}]{2014..stepanovs..a} 
D. Stepanovs, Ch. Fendt, \apj, \textbf{793}, 31 (2014).

\bibitem[\protect\citeauthoryear{Stepanovs, Fendt and Sheikhnezami}{2014}]{2014..stepanovs..b} 
D. Stepanovs, Ch. Fendt, S. Sheikhnezami, \apj, \textbf{796}, 29 (2014).

\bibitem[\protect\citeauthoryear{Stephens et al.}{2014}]{Stephens+14} 
I. W. Stephens, L. W. Looney, W. Kwon, M. Fern\'andez-L\'opez, A. M.
Hughest, L. G. Mundy, R. M. Crutcher, Z.-Y. Li, R. Rao, Nature,
\textbf{514}, 597 (2014)

\bibitem[\protect\citeauthoryear{Stone and Norman}{1992}]{1992ApJ...389..297S}
J.M. Stone, M.L. Norman, \apj, \textbf{389}, 297 (1992)

\bibitem[\protect\citeauthoryear{Stone and Norman}{1994}]{1994ApJ...433..746S}
J.M. Stone, M.L. Norman, \apj, \textbf{433}, 746 (1994).

\bibitem[\protect\citeauthoryear{Stone and Hardee}{2000}]{2000ApJ...540..192S}
J.M. Stone, P.E. Hardee, \apj, \textbf{540}, 192 (2000).

\bibitem[\protect\citeauthoryear{Suzuki and Inutsuka}{2014}]{Suzuki14}
T. K. Suzuki and S.-I. Inutsuka, \apj, \textbf{784}, 121

\bibitem[\protect\citeauthoryear{Tavecchio et al.}{2000}]{Tavecchio+00} 
F. Tavecchio, L. Maraschi,  R. M. Sambruna, C. M. Urry, \apj, 
\textbf{544}, L23 (2000)

\bibitem[\protect\citeauthoryear{Taylor and
Zavala}{2010}]{TaylorZavala2010} G. B. Taylor, R. Zavala,
\apj, \textbf{722}, L183 (2010)

\bibitem[\protect\citeauthoryear{Tchekhovskoy et al.}{2008}]{Tchekhovskoy-08}
A. Tchekhovskoy, J.C. McKinney, R. Narayan, \mnras, \textbf{388}, 551 (2008)

\bibitem[\protect\citeauthoryear{Tchekhovskoy, McKinney and
Narayan}{Tchekhovskoy et al.}{2009}]{Tchekhovskoy-09} A. Tchekhovskoy,
J.C. McKinney, R. Narayan, \apj, \textbf{699}, 1789 (2009)

\bibitem[\protect\citeauthoryear{Tchekhovskoy, Narayan and
McKinney}{Tchekhovskoy et al.}{2010}]{2010NewA...15..749T} A. Tchekhovskoy,
R. Narayan, J.C. McKinney, New Astronomy, \textbf{15}, 749 (2010)

\bibitem[\protect\citeauthoryear{Tchekhovskoy et
al.}{2010}]{2010ApJ...711...50T} A. Tchekhovskoy,
R. Narayan, J.C. McKinney, \apj, \textbf{711}, 50 (2010)

\bibitem[\protect\citeauthoryear{Tchekhovskoy, Narayan and
McKinney}{Tchekhovskoy et al.}{2011}]{Tchekhovskoy-11} A.
Tchekhovskoy, R. Narayan, J.C. McKinney,
\mnras, \textbf{418}, L79 (2011)

\bibitem[\protect\citeauthoryear{Tchekhovskoy, McKinney and
Narayan}{Tchekhovskoy et al.}{2012a}]{Tchekhovskoy-2012a} A.
Tchekhovskoy, J.C. McKinney,
R. Narayan. J. Phys. Conference Series, \textbf{372}, 012040, 
arXiv:1202.2864 (2012)

\bibitem[\protect\citeauthoryear{Tchekhovskoy and
    Giannios}{2014}]{2014arXiv1409.4414T} A. Tchekhovskoy,
D. Giannios, \mnras, \textbf{447}, 327 (2014)

\bibitem[\protect\citeauthoryear{Tchekhovskoy et
al.}{2014}]{2014MNRAS.437.2744T} A. Tchekhovskoy,
B.D. Metzger,  D. Giannios, L.Z. Kelley,  \mnras, \textbf{437},
2744 (2014)


\bibitem[\protect\citeauthoryear{Tchekhovskoy} 
{2015}]{Tchekhovskoy-15} A. Tchekhovskoy,
in \textit{Formation and Destruction of Active Galactic Nuclei
Jets}, Contopoulos et al. (eds), in press Springer (2015)


\bibitem[\protect\citeauthoryear{Todo et al.}{1993}]{1993ApJ...403..164T}
Y. Todo, Y. Uchida, T. Sato, R. Rosner, \apj, \textbf{403}, 164 (1993).

\bibitem[\protect\citeauthoryear{Tregillis, Jones and
Ryu}{2004}]{Tregillis+04} I. L. Tregillis, T. W. Jones, D. Ryu,
\apj, \textbf{601}, 778 (2004)

\bibitem[\protect\citeauthoryear{Tzeferacos et al.}{2009}]{2009MNRAS.400..820T}
P. Tzeferacos, A. Ferrari, A. Mignone, C. Zanni, G. Bodo, S.
Massaglia, \mnras, \textbf{400}, 820 (2009).

\bibitem[\protect\citeauthoryear{Tzeferacos et al.}{2013}]{2013MNRAS.428.3151T}
P. Tzeferacos, A. Ferrari, A. Mignone, C. Zanni, G. Bodo, S.
Massaglia, \mnras, \textbf{428}, 3151 (2013).

\bibitem[\protect\citeauthoryear{Uchida and Shibata}{1984}]{US84} 
Y. Uchida, K. Shibata, \pasj, \textbf{36}, 105 (1984)

\bibitem[\protect\citeauthoryear{Uchida and Shibata}{1985}]{US85}
Y. Uchida,  K. Shibata, \pasj, \textbf{37}, 515 (1985)

\bibitem[\protect\citeauthoryear{Ustyugova et al.}{1999}]{1999ApJ...516..221U}
G.V. Ustyugova, A.V. Koldoba, M.M. Romanova, V.M. Chechetkin, R.V.E.
Lovelace, \apj \textbf{516}, 221 (1999).

\bibitem[\protect\citeauthoryear{Ustyugova et al.}{1995}]{1995ApJ...439L..39U}
G.V. Ustyugova, A.V. Koldoba, M.M. Romanova, V.M. Chechetkin, R.V.E.
Lovelace, \apj, \textbf{439}, 39 (1995).

\bibitem[\protect\citeauthoryear{Vaidya et al.}{2011}]{2011ApJ...742...56V}
B. Vaidya, Ch. Fendt, H. Beuther, O. Porth, \apj, \textbf{742}, 56 (2011).

\bibitem[\protect\citeauthoryear{van-Ballegooijen}{1989}]{vanB89}
A. A. van-Ballegooijen, in \textit{Accretion Disks and Magnetic
Fields in Astrophysics}, ed. G. Belvedere, 99 (1989)

\bibitem[\protect\citeauthoryear{von Rekowski and
Brandenburg}{2004}]{2004A&A...420...17V} B. von Rekowski, A.
Brandenburg, \aap, \textbf{420}, 17 (2004).

\bibitem[\protect\citeauthoryear{von Rekowski et
al.}{2003}]{2003A&A...398..825V} B. von Rekowski, A. Brandenburg,
W. Dobler, \aap, \textbf{398}, 825 (2003).

\bibitem[\protect\citeauthoryear{Wardle et al.}{1998}]{Wardle+98}
J. F. C. Wardle, D. C. Homan, R.  Ojha, D. H. Roberts,  \nat,
\textbf{395}, 457 (1998)

\bibitem[\protect\citeauthoryear{Wardle and
K\"onigl}{1993}]{1993ApJ...410..218W} M. Wardle, A. K\"onigl, \apj,
\textbf{410}, 218 (1993).

\bibitem[\protect\citeauthoryear{Winters, Balbus and
Hawley}{2003}]{Winters03} W. F. Winters, S. A. Balbus, J. F. Hawley, 
\mnras, \textbf{340}, 519 (2003)

\bibitem[\protect\citeauthoryear{Wykes et al.}{2013}]{Wykes+13} 
S. Wykes, et al., \aap, \textbf{558}, A19 (2013)

\bibitem[\protect\citeauthoryear{Zamaninasab et
al.}{2014}]{2014Natur.510..126Z} M. Zamaninasab, E. Clausen-Brown,
T. Savolainen, A. Tchekhovskoy, \nat, \textbf{510}, 126
(2014)

\bibitem[\protect\citeauthoryear{Zanni et al.}{2007}]{2007A&A...469..811Z}
C. Zanni, A. Ferrari, R. Rosner, G. Bodo, S. Massaglia, \aap,
\textbf{469}, 811 (2007).

\bibitem[\protect\citeauthoryear{Zdziarski et
al.}{2014}]{2014arXiv1410.7310Z} A.A. Zdziarski, M. Sikora,
P. Pjanka, A. Tchekhovskoy, arXiv:1410.7310 (2014)

\bibitem[\protect\citeauthoryear{Zinnecker et al.}{1998}]{1998Natur.394..862Z}
H. Zinnecker, M.J. McCaughrean, J.T. Rayner, \nat, \textbf{394}, 862 (1998).

% 

\end{thebibliography}
\end{document}